\documentclass[12pt]{article}
\usepackage[utf8]{inputenc}
\usepackage{braket}
\usepackage[english]{babel}
\usepackage[utf8]{inputenc}
\usepackage{siunitx}
\usepackage{bm}
\usepackage{graphicx, caption, subcaption}
\usepackage{verbatim}
\usepackage{amsmath}
\usepackage{amssymb}
\newcommand{\bbGamma}{{\mathpalette\makebbGamma\relax}}
\newcommand{\makebbGamma}[2]{%
  \raisebox{\depth}{\scalebox{1}[-1]{$\mathsurround=0pt#1\mathbb{L}$}}%
}
\DeclareMathAlphabet{\mymathbb}{U}{BOONDOX-ds}{m}{n}

\begin{document}

\title{Simulating a Catalyst induced Quantum Dynamical Phase Transition of a Heyrovsky reaction with different models for the environment.}
\author{Fabricio S. Lozano-Negro, Marcos A. Ferreyra-Ortega,  Denise Bendersky,\\ Lucas  Fern\'{a}ndez-Alc\'{a}zar and Horacio M. Pastawski}
%\address{Instituto de F\'{\i}sica Enrique Gaviola and Facultad de Matem\'{a}tica, Astronom\'{\i}a, F\'{\i}sica y Computaci\'{o}n. Universidad Nacional de C\'{o}rdoba. 5000 C\'{o}rdoba, Argentina.}
\maketitle
\begin{abstract}
Through an appropriate election of the molecular orbital basis, we show analytically that the molecular dissociation occurring in a Heyrovsky reaction can be interpreted as a Quantum Dynamical Phase Transition, i.e., an analytical discontinuity in the molecular energy spectrum induced by the catalyst. The metallic substrate plays the role of an environment that produces an energy uncertainty on the adatom. This broadening induces a critical behavior not possible in a quantum closed system. We use suitable approximations on symmetry, together with both  Lanczos and canonical transformations, to give analytical estimates for the critical parameters of molecular dissociation. This occurs when the bonding to the surface is \(\sqrt{2}\) times the molecular bonding. This value is slightly weakened for less symmetric situations. However simple, this conclusion involves a high order perturbative solution of the molecule-catalyst system. This model is further simplified to discuss how an environment-induced critical phenomenon can be evaluated through an idealized perturbative tunneling microscopy set-up. In this case, the energy uncertainties in one or both atoms are either Lorentzian or Gaussian. The former results from the Fermi Golden Rule, i.e., a Markovian approximation. The Gaussian uncertainty, associated with non-Markovian decoherent processes, requires the introduction of a particular model of a spin bath. The partially coherent tunneling current is obtained from the Generalized Landauer-B\"uttiker Equations. The resonances observed in these transport parameters reflect, in many cases, the critical properties of the resonances in the molecular spectrum. 
\end{abstract}

\section{Introduction.}

\subsection{When does a molecule break apart?}
One of  unsolved mysteries of Chemistry  \cite{Ba11} is how do molecules form and dissociate. This has closely connected variants: Is there a clear distinction between two individual atoms and a new molecular species? When could a solution of the Schr\"{o}dinger equation as a function of the distance between the atom $A$ and the ion $B$, as in $\mathrm{H}_{2}^{+}$, describe an actual bond? Indeed, the deep roots of these questions can be traced back to Quantum Mechanics itself. In fact, the typical algebra of a quantum Chemistry computation does not yield any sort of discontinuity that could discriminate between both cases. Of course, in a lecture one can convincingly justify the minimum of the bonding energy with distance and everyone seems enough comfortable with it. However, one might hint that some subtle ingredient is still missing, perhaps some sort discontinuity of those which are quite common in thermodynamics or other areas of Physics. Nevertheless, this does not show up at this simple level of description. If one actually starts from an atom and an ion approaching from far away, a quantum dynamical calculation would not give access to this energy minimum. This would require energy relaxation through some coupling with external degrees of freedom that are not contained in the model we usually present to our students.

\subsection{More is different.}
The environment is the under-covered protagonist of  an already legendary paper\cite{An61} by P. W. Anderson. He discussed an inspiring phenomenon that guides our search: the oscillatory dynamics of inversion of the Ammonia pyramidal molecule between its two equivalent ``umbrella configurations'', $A$ and $B$. Its ground state is no other than the bonding symmetric state, which satisfies the theorem that precludes the existence of a dipolar moment. However, if the molecule is initially set on one of its unsymmetrical states, it would periodically swap between them. This oscillation, with a frequency of about $2.4\times10^{10}\mathrm{Hz}$, enabled the design of a maser\cite{FeynVol3}. Since the phenomenon involves quantum tunneling, one might immediately realize that this rate depends on the mass. Indeed, Nature repeats herself with the heavier ion $\mathrm{P}$ replacing $\mathrm{N}$ as in Hydrogen phosphide, $\mathrm{PH}_{3}$, this time with a much longer period. In contrast, the heavier Phosphorus trifluoride, $\mathrm{PF}_{3},$ has never been seen to invert by itself in spite that a calculation would yield an experimentally accessible time. Thus, as the mass is increased some {\it spontaneous symmetry violation} occurs.  How is this described? Anderson, does not give an explicit answer, but he reaffirms the situation mentioning that for a sugar molecule, with about 40 atoms, one does not expect a similar tunneling. Then, he introduced the concept of \textit{emergent phenomena} by describing that the increase of the atomic mass, molecular size or number of atoms would yield a new limit, not necessarily contained in the mathematical solution previously available. Indeed, this idea, applied to Standard Model of elementary particles, gave raise to the prediction, hinted by Anderson, of the Higgs boson\cite{Hig14}.

The objective of the present paper is to review  the above situation under the light of Quantum Mechanics of open systems. In Section 2, we adopt the simplest model that enables energy relaxation and embodies this discontinuity phenomenon. This allows us to describe and quantify the Heyrovsky reaction as critical phenomenon enabled by the thermodynamic limit, much of the type inferred by Anderson. In Section 3, the dimer-environment system is further simplified to allow the search for environment induced critical phenomena that could yield measurable spectral signatures that might appear in tunneling spectroscopy set-up. For this we consider the current for different environments and electrode configurations with the Keldysh linearized equations. 

\section{Towards a model of Catalytic Molecular Dissociation.}

\subsection{An inspiring SWAP-gate experiment.}

In view of the discussed ideas it was very exciting when, trying to optimize a quantum SWAP gate in a solid-state NMR experiment, a team at C\'{o}rdoba that included one of the authors observed a non-analytic transition in the swapping dynamics. A pair of interacting nuclear spins  $A$ and $B$ whose Rabi oscillation had to be quenched when the polarization maximized at one nucleus, switched from a damped oscillation\cite{KuBEr74} to an over-damped dynamics\cite{LUPa98,ALPa07}. These experiments made it clear that the origin of the spontaneous symmetry violation had to be traced back to the infinite number of degrees of freedom of the environment, i.e. the unlimited cloud of spins that surrounds spin $B$, and whose effect is condensed in the Fermi Golden Rule (FGR) rate $\Gamma_{B}$, which also can be seen as an energy broadening \cite{RfPa06}. Indeed, by rotating the crystal, the ratio between the SWAP frequency, $2V_{AB}/\hbar$, and the coupling with the environment, $\Gamma_{B}$, could be varied and the oscillation could be turned on and off at will. In that approximation, the two-spin system with such a spin bath can be described by an effective \textit{non-Hermitian} Hamiltonian that results from the partitioning formalism as initially introduced by P. O. L\"{o}wdin in Molecular Physics and H. Feshbach in Nuclear Physics \cite{Low68, Fe58, 
Ro09}:

\begin{equation}\label{eq:non-hermitian-H}
\widetilde{\mathbb{H}}\mathbb{=}\left[
\begin{array}
[c]{cc}
E_{A} & -V_{AB}\\
-V_{AB} & ~~~~~~E_{B}-\mathrm{i}\Gamma_{B}%
\end{array}
\right],
\end{equation}
with $E_{A}=E_{B}=E_{0}.$ Thus, the complex
eigenvalues
\begin{equation}
\widetilde{E}_{a/b}=\left\{
\begin{array}
[c]{ccc}%
E_{0}-\mathrm{i}\tfrac{1}{2}\Gamma_{B}\pm\sqrt{\left\vert V_{AB}\right\vert
^{2}-\left(  \dfrac{\Gamma_{B}}{2}\right)  ^{2}} & \text{for} & \left(
\dfrac{\Gamma_{B}}{2}\right)  ^{2}\leq\left\vert V_{AB}\right\vert ^{2}\\
E_{0}-\mathrm{i}\tfrac{1}{2}\Gamma_{B}\pm\mathrm{i}\sqrt{\left(  \dfrac{\Gamma_{B}}{2}\right)  ^{2}-\left\vert V_{AB}\right\vert ^{2}} & \text{for} & \left(
\dfrac{\Gamma_{B}}{2}\right)  ^{2}>\left\vert V_{AB}\right\vert ^{2}%
\end{array}
\right.
\end{equation}
show that the complex bonding and antibonding energies, $\widetilde{E}_{b}$ and $\widetilde{E}_{a}$, drift in the complex plane from the real axis towards the imaginary axis as $\Gamma_{B}/2V_{AB}$ is increased from zero. Similar energy shifts and broadenings result when one accounts for the escape toward electrodes\cite{DAPa90,Pa91} or consider many-body interactions such as electron-electron within a Random Phase Approximation \cite{DALP07} or electron-phonon interactions in a local phonon approximation\cite{Pa92,PaFoMe02}.

\begin{figure}
    \centering
    \includegraphics[width=0.9\textwidth]{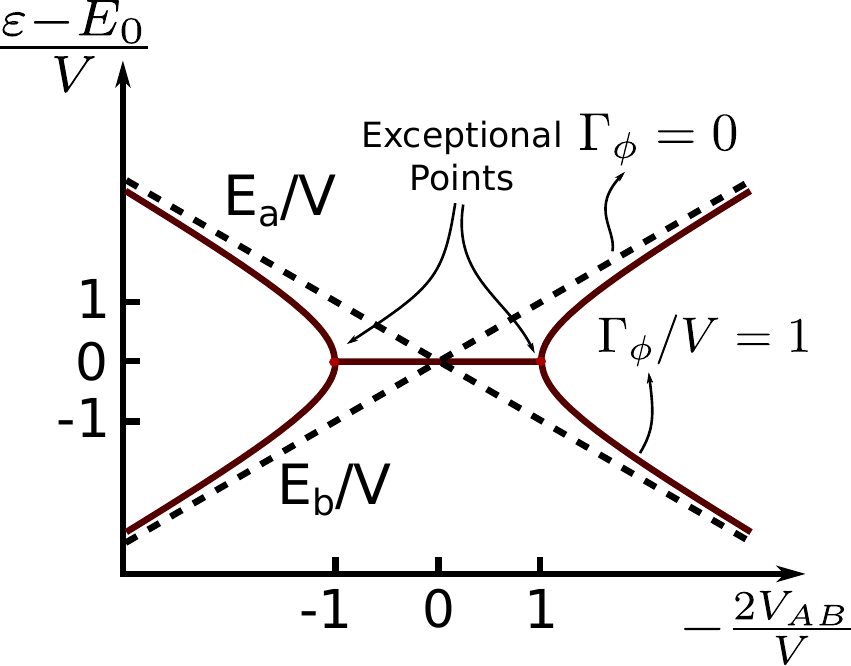}
    \caption{Real part of the eigenvalues of Hamiltonian \eqref{eq:non-hermitian-H} in absence of a environment (Hermitian spectrum, dashed lines) and in presence of an environment (Non-Hermitian spectrum, solid lines).}
    \label{fig:00}
\end{figure}

It is evident that if $V_{AB}$ decreases, the spectrum has a critical point where the real components of the eigenvalues, the \textit{bonding and antibonding energies}, collapse and stay degenerate. Simultaneously, the imaginary components, associated with \textit{energy broadenings} or  \textit{decay rates}, bifurcate. This, in turn, induces in a bifurcation in the \textit{complex frequency spectrum} that could manifest in the dynamics of actual observables \cite{EGM20}. These observables might be described by a Keldysh Quantum Fields formalism \cite{DaPaA05,DALP07} or by a Liouville equation \cite{KuBEr74}. The phenomenon was then dubbed Quantum Dynamical Phase Transition (QDPT) because the observables that result from the quantum dynamics are non-analytic functions of the control parameter $\Gamma_{B}/2V_{AB}$. This phenomenon has a natural analogy with the transition towards the over-damped regime of a classical harmonic oscillator. It also shares some common features with the super-radiance and superconducting  transitions \cite{C+CoRa05},  the synchronization of clocks described by C. Huygens, the tuning of piano strings \cite{We77} and the collective phase transitions described for non-linear systems \cite{Stro12}. Indeed, a number of classical oscillatory systems have also shown exceptional points in the spectrum and their associated dynamical phase transition \cite{An54, CeK09, RoBi15, Do_Ko20} and they have been exploited in a wide range of innovative applications that include, the sub-wave length field control  \cite{Bu_Pa14}, energy harvesting and energy management \cite{LFAESK19, FALEAK20, FAKK21}, among many others. All those cases contain, often hidden behind FGR-like approximations, the thermodynamic limit required by Anderson.

In this context, it is inspiring to observe a dynamical simulation of a Heyrovsky reaction\cite{Sa_Sc11}. As an $\rm{H}_2$ reaches a critical distance to the catalyst,  a sudden release of the farthest $\mathrm{H}$ occurs while the other $\mathrm{H}$ remains bonded to the metallic substrate. This suggested that molecular dynamics simulations might contain the essential ingredients for the molecule dissociation (or formation) as function of some control parameter. This sets our goal to describe the catalytic molecular dissociation as QDPT with a minimal modeling. We may paraphrase R. B. Woodward and R. Hoffmann: \textit{The lack of numbers in our discussion is not a weakness, it is its greatest strength. Precise numerical values would have to result from some specific sequence of approximations. But an argument from first principles or symmetry, of necessity qualitative, is in fact much stronger than the deceptively authoritative numerical result. For, if the simple argument is true, then any approximate method, as well as the now inaccessible exact solution, must obey it\cite{WoHo69}.}

A first \textit{naive} approach to molecular dissociation was a direct adaptation from what we learned from systems of spins. Indeed, in that case the spins can be formally mapped to a tight-binding spinless electronic model through the Jordan-Wigner transformation  \cite{JW28, DaPaA05, RfPa06}. 
A \textit{spin up} corresponds with an \textit{electron} occupying the orbital state, while \textit{spin down} is identified with its absences, a \textit{hole}. The flip-flop interactions are equivalent to the tunneling of electrons enabled by the overlap integral in an extended H\"{u}ckel model. Thus, the SWAP gate where one of the spins interacts with a chain of spins at high temperature becomes a promising variant of the Anderson-Newns model\cite{An61,Nw69,DeBMPa08,SaKSc08}.
Indeed, we are left with a dimer $AB$ approaching the surface of a catalyst with an \textbf{on-top} configuration as shown on Fig. 2. Thus, when atom $B$ gets at a distance $R$ from the first atom in the metal surface with a single $d_{{\large z}^{{\large 2}}}$ orbital with a matrix element $V_{0}$. We may safely neglect the trivial spin indexes and the vector notation of the spin-orbital states in favor of a more transparent view. 
In this simplified notation, the Hamiltonians of the molecule, $\hat{H}_{AB}$, and the substrate, $\hat{H}_{\mathcal{E}},$ as well as their coupling, $\hat{V}_{AB-\mathcal{E}}$, are expressed in terms of electron creation and annihilation operators:

\begin{equation}
\hat{H}_{AB}=(E_{A}\hat{c}_{A}^{+}\hat{c}_{A}^{{}}+E_{B}\hat{c}_{B}^{+}\hat
{c}_{B}^{{}})+V_{AB}(\hat{c}_{A}^{+}\hat{c}_{B}^{{}}+\hat{c}_{A}^{{}}\hat
{c}_{B}^{+}),
\end{equation}

\begin{equation}
\hat{H}_{\mathcal{E}}=\sum_{k}E_{k}\hat{c}_{k}^{+}\hat{c}_{k}^{{}},
\end{equation}\begin{equation}
\hat{V}_{AB-\mathcal{E}.}=V_{0}(\hat{c}_{B}^{+}\hat{c}_{1}^{{}}+\hat{c}_{1}^{+}\hat{c}_{B}^{{}}).
\end{equation}
These operators act on the \textit{Fermi sea,} $\left\vert F\right\rangle =\Pi _{k=0}^{k_{F}}\hat{c}_{k}^{+}\left\vert 0\right\rangle ,$ that describes the catalyst \(d\)-band. Of course, a local orbital $d_{{\Large z}^{2}}$ of the atom at the surface lattice site $R_{1},$ that is $\varphi_{1}(r)=\left\langle r\right\vert \hat{c}_{1}^{+}\left\vert 0\right\rangle ,$ is expressed as an adequate Bloch sum over the states $k$ of the metallic substrate $\hat{c}_{1}=\sum_{k}\exp[-\mathrm{i}R_{1}k]\hat{c}_{k}.$ In an adsorbed molecule, a non-Hermitian effective Hamiltonian results when an atomic energy $E_{B}$  is shifted by self-energy correction $\Sigma_{B}(\varepsilon)$ evaluated up to second order perturbation of each of \(N\) substrate \(k\) states. As one takes \(N \rightarrow \infty\)  

\begin{equation}\label{eq:2nd-orderSigma}
\Sigma_{B}(E_{B})=\lim_{\eta\rightarrow0}\lim_{N\rightarrow\infty}\sum_{k}\frac{\left\vert
V_{Bk}\right\vert ^{2}}{E_{B}-E_{k}+\mathrm{i}\eta}=\Delta_{B}%
-\mathrm{i}\Gamma_{B}%
\end{equation}
resulting in the Fermi Golden Rule exponential decay rate:

\begin{equation}
\Gamma_{B}=\pi\left\vert V_{0}\right\vert ^{2}N_{1}.
\end{equation}

Here, $N_{1}$ is the local density of states at the surface orbital  $d_{{\large z}^{\large 2}}$ evaluated at energy $E_{B}$. This apparently simple trick is dubbed \textit{wide band approximation} (WBA).  In our case, the WBA is not precise enough, and we develop a more detailed description of the dynamics \cite{RfPa06, Ga_Or21} where the dependence of $\Sigma_{B}(\varepsilon)$ on the energy $\varepsilon$ is included. For that, one could  resort to the Dyson equation for the retarded Green's function as briefly described in Appendix A. However, for the moment we stay within the WBA.

 Hence, the non-Hermitian Hamiltonian operator looks:\begin{equation}
\hat{\widetilde{H}}_{AB}=E_{A}\hat{c}_{A}^{+}\hat{c}_{A}^{ }%
+[E_{B}+\Delta_{B}-\mathrm{i}\Gamma_{B}]\hat{c}_{B}^{+}\hat{c}_{B}^{{}}%
+V_{AB}(\hat{c}_{A}^{+}\hat{c}_{B}^{{}}+\hat{c}_{B}^{+}\hat{c}_{A}^{{}%
}).\label{H_AB-eff}%
\end{equation}
The exact complex eigenvalues are obtained from a non-linear equation that yields a Self-Consistent Fermi Golden Rule. Thus, our simplified problem becomes
\begin{equation}
\widetilde{\mathbb{H}}_{AB}\overrightarrow{u}=\varepsilon\overrightarrow{u},
\end{equation}
leading to the equation for the eigenvalues
\begin{equation}
\left[  \varepsilon-E_{A}\right]  \left[  \varepsilon-E_{B}-\Sigma_{B}\right]
-\left\vert V_{AB}\right\vert ^{2}=0,
\end{equation}
which coincide with the \textit{poles} of the \textit{retarded} Green's function \(\mathbb{G}(\varepsilon)=1/[\varepsilon\mathbb{I}-\widetilde{\mathbb{H}}_{AB}]\). If we consider that $\left\vert V_{0}\right\vert ^{2}=0$ corresponds to the isolated molecule, we see that as $V_{0}$ increases there is a critical value when the line-broadening matches the level splitting, i.e.
\begin{align}
\Gamma_{B}^\mathrm{crit.}  & =\pi  \left\vert V_{0}^\mathrm{crit.}\right\vert ^{2} \times  N_{1} =2V_{AB},\\
\end{align}Indeed, one might conceive a situation where the orbital $A$ is also under the action of a similar environment and it has its own energy uncertainty, $\Gamma_{A}$. In this case, the condition for a QDPT is $\left\vert \Gamma
_{A}-\Gamma_{B}\right\vert \geq2V_{AB}$. Although, identical environments can yield a QDPT in the observable frequency spectrum $\omega$, this result requires a formalism that ensures the conservation of the electronic density. This is provided by the complete system of Keldysh equations of the quantum fields. These should be solved at least in their linearized version provided by  the Generalized Landauer-B\"{u}ttiker Equations (GLBE) \cite{Pa92}. Alternatively, one might resort to the simpler master equation for the density matrix, in the density conserving approximations\cite{CPz17} developed  by Lindblad and independently by Gorini, Kossakowski and Sudarshan, in any of their appropriate variants \cite{DaPaA05,Pa07,ALPa07}. 

Interestingly, we see that, despite of particular details, the different environments such as electrodes, voltage probes, a metallic catalyst, a thermal bath oscillators, or the solvent degrees of freedom, play a similar role in what concerns to energy uncertainties Refs.\cite{PaFoMe02,SaKSc06,CaBMPa10,CaFe+Pa14}. For example, to represent solvent states one may use a set of oscillators
\begin{align*}
\hat{H}_{solv.} &  =\sum_{q}\hbar\omega_{q}\left(  \hat{b}_{q}^{+}\hat{b}%
_{q}+\tfrac{1}{2}\right)  \\
\hat{V}_{B-solv.} &  =\sum_{q}V_{g}\hat{c}_{B}^{+}\hat{c}_{B}\left(  \hat
{b}_{q}^{+}+\hat{b}_{q}\right).
\end{align*}
where the interaction results in a polaronic model and a contribution to the level broadening $\Gamma_{B}$. Within suitable approximations as the Debye model for the oscillator bath, $\Gamma_{B}$ results proportional to the temperature Refs.\cite{Pa92,SaKSc06,CaBMPa10}. 

The crucial drawback that prevents the direct application of such model to a catalytic dissociation is that the weak overlap with a broad $s$-band is not strong enough to induce a QDPT. Thus, it is clear the phenomenon should be sought by incorporating the strongly directional orbitals of a $d$-band. Still, the analysis requires a somewhat different approach as the molecular bonding exceeds the band width.

\subsection{The Catalytic Heyrovsky reaction as a QDPT.}

We represent a diatomic homonuclear molecule ($A$-$B$) using the covalent gain per electron $V_{AB}$. As $A$-$B$ approximates the catalyst, the orbital \(B\) and the surface \(d_z\) pointing towards develop a coupling $V_{0}$, associated with the overlap integral. The substrate electron states relevant for  catalysis are in half-filled $d$ band of width $4V_d$.
 Since the molecular bonding energy exceeds the width of the $d$-band, $V_{AB}>4V_d,$ one can not directly invoke the simplest $2\times2$ non-Hermitian model discussed initially. However, we might invoke Hoffmann's rule: \textit{metal-adsorbate bonding is accomplished at the expense of bonding within the metal and the adsorbed molecule}\cite{Ho88}. We will make a specific elaboration of this concept, that with a few mathematical subtleties, will disclose the QDPT.

\begin{figure}
    \centering
    \includegraphics[width=0.5\textwidth]{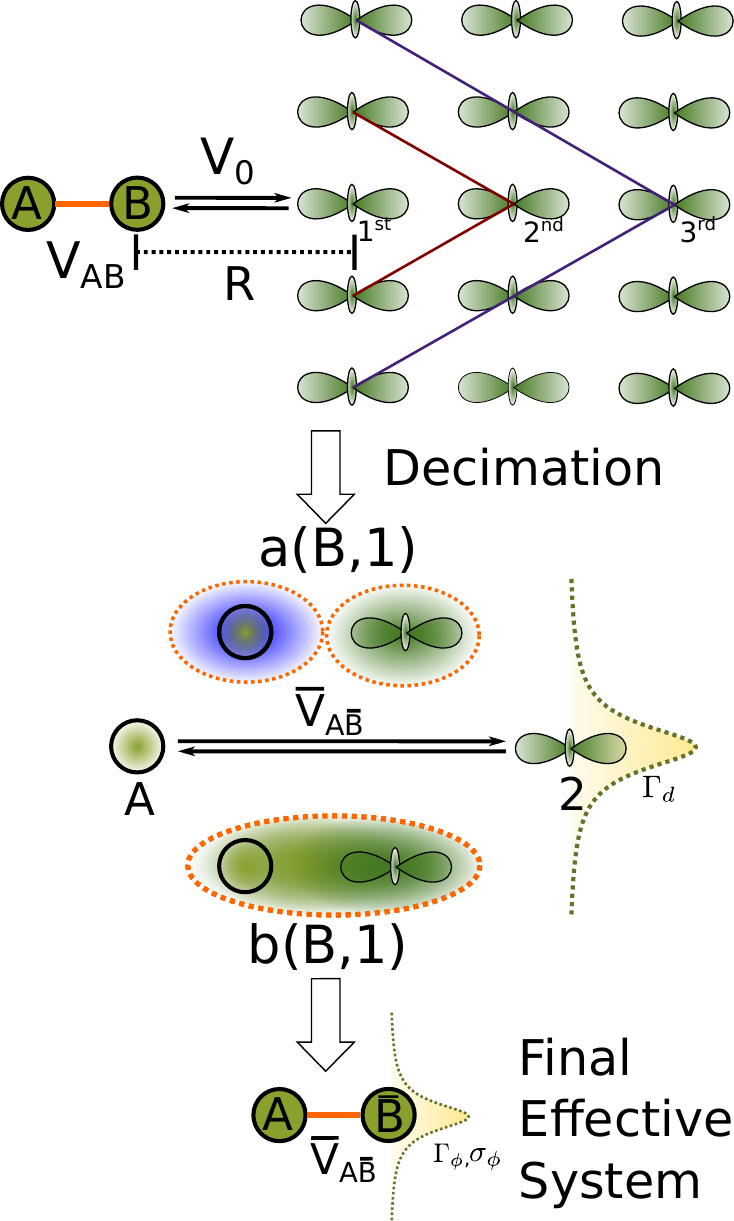}
    \caption{Schematic representation of the transformation of a molecule approaching (perpendicularly) a metallic surface. (1) The substrate orbitals are redefined using Lanczos unitary transformations which combines orbitals with increasing coordination order. (2) The decimation of all orbitals beyond the second order introduces an uncertainty in the energy of the effective orbital of order two of about the full \(d\)-bandwidth ($\Gamma_d$), while the strong coupling between the $B$ orbital and the 1st surface \(d_{z^2}\) is represented by their bonding and antibonding states. (3) Final model where the orbital \(A\) has now a through-bond coupling with the collective orbital of 2nd order, now renamed \(\bar{B}\). The energy broadening of \(\bar{B}\)  is either Lorentzian ($\Gamma_{\phi}$) or Gaussian ($\sigma_{\phi}$).}
    \label{fig:01}
\end{figure}

Without much loss of generality we might consider all the energies, $E_{A}$, $E_{B}$, $E_{1}$, and $E_{2}$ all equal to $E_{d}$ ( which eventually could be zero). That is, we consider a precise resonance between the atomic energies and the \(d\)-band. This restriction will be relaxed in Appendix \ref{A:Asymmetric} in a more detailed discussion of an asymmetric case. For our main argument it is enough to assume a monotonic increasing of the bonding strength $V_{0}$ between the atom $B$ and the catalyst's surface orbital as the molecule approaches the surface.

The isolated molecule would have the bonding energy $\varepsilon_{b}=-V_{AB}$ and antibonding one $\varepsilon_{a}=+V_{AB}$, that corresponds to the symmetric and anti-symmetric combination of orbitals.  By creating an electron in orbital $A$  or $B$, in addition to the Fermi sea $\vert F\rangle $, we arrive at usual notation $\left\vert A\right\rangle =\hat{c}_{A}^{+}\left\vert F\right\rangle $ and $\left\vert B\right\rangle=\hat{c}_{B}^{+}\left\vert F\right\rangle $). Thus, the molecule bonding state is $\left\vert b(AB)\right\rangle =\left(  \left\vert A\right\rangle +\left\vert B\right\rangle \right)  /\sqrt{2}$ and the anti-bonding one is $\left\vert a(AB)\right\rangle =$ $\left(  \left\vert A\right\rangle -\left\vert B\right\rangle \right)  /\sqrt{2}$.  This energy structure of the molecule is tested through two tunneling probes that are attached to atom $A$ (through-atom tunneling), or one at atom $A$ and the other at $B$ (through-bond tunneling). Since  $V_{AB}>4V_d$, the energies of these molecular states are outside the $d$-band.

The portion of the electronic structure of the metal that participates in the catalysis can be described according to its projection on the metallic surface orbital. It simple structure is revealed by the orthogonal basis obtained from a  Lanczos tridiagonalization procedure. The Physics behind this method was discussed in detail in a number of papers by Haydock, Heine and Kelly \cite{HaHeKe72, HaHeKe75}. The key physical strength of this basis is that the local properties of an ordered or disordered crystal surface\cite{ DesSp96} are represented in terms of a tight-binding linear chain. Each new \textit{orbital} is built as a linear combination of the atomic orbitals at a coordination sphere of a given order of neighbors (upper panel Fig. 2 shows those of 2nd and 3rd order) conveniently orthogonalized to the previous ones. The diagonal terms soon converge to $E_{d}$ and the non-diagonal terms, to the hopping $V_{d}=\Gamma_{d}$ that determines the bandwidth $4V_{d}$ of the $d$-band.

As the molecule approaches the surface, the coupling $V_{0}$ increases and the molecular bonding state becomes $\left\vert b(B1)\right\rangle$, the asymptotic bonding combination between the $\left\vert B\right\rangle$ orbital and $\left\vert1\right\rangle$, the $d_{z^{2}}$ orbital at the surface atom. Similarly, the molecular antibonding state progressively becomes the antibonding combination $\left\vert a(B1)\right\rangle $, i.e. in the extreme, it becomes an isolated covalent bond. Under this condition, the atomic orbital $A$ and the symmetric combination of the $d$-orbitals in the second shell of neighbors acquire a effective \textit{positive through-bond coupling}:
\begin{equation}
\bar{V}_{A\bar{B}}\equiv V_{A2}\simeq2\left(  {\frac{1}{\sqrt 2}}V_{12}\right)\dfrac{1}{V_{0}}\left(  {\frac{1}{\sqrt 2}}V_{AB}\right)=\dfrac{V_d\times V_{AB}}{V_{0}}\gtrsim\dfrac{2V_d^{2}}{V_{0}},
\label{Eq13}
\end{equation}
leading to $$\left\vert b(A2)\right\rangle=\left(  \left\vert A\right\rangle -\left\vert 2\right\rangle \right)  /\sqrt{2} $$ and $$\left\vert a(A2)\right\rangle =\left(  \left\vert A\right\rangle +\left\vert 2\right\rangle \right)  /\sqrt{2}$$ respectively. This secondary through-bond coupling\cite{LePaD90}, absent in the original basis, singles out a collective molecular orbital $\left\vert 2\right\rangle$ that includes a coordination sphere of second neighbors of the $B$ atom from the rest of the $d_{z}$-band.  Indeed, this would be the eigenstates if the further layers of the catalyst were ignored. However, according to the Lanczos basis, the collective orbital is still renormalized by \(\Sigma_{d}(\varepsilon)\simeq  (\varepsilon -E_{d})/2 -\mathrm{i}V_d\) that accounts for the rest of the infinite substrate. The resulting bonding and antibonding combinations,  manifests as \textit{two resonances within the $d$-band}. However, their energy distance will be diminished from $2{V}_{A2}$ by the action of the uncertainty $\Gamma_{d}=V_d$ induced by the $d$-band. Under this condition, a further approach would increase $V_{0}$ weakening the through-bond coupling ${V}_{A2}$ and hence the bonding and antibonding character of the resonances within the $d$ band. The vanishing square root, which regularly controls the energy splitting, induces the collapse of the resonant energies $\tilde {E}_{b(A2)}$ and $\tilde{E}_{a(A2)}$, which are the roots of the effective determinant:
\begin{equation}
\left[  \varepsilon-E_{A}\right]  \left[  \varepsilon-E_{2}-\Sigma_{d}(\varepsilon)\right]
-\left\vert V_{A2}\right\vert ^{2}=0.
\end{equation}
The discriminant vanishes at
\begin{equation}
V_{0}^{\mathrm{crit.}}=\sqrt{2}\dfrac{V_d\times V_{AB}}{V_d}\simeq \sqrt{2} V_{AB},
\end{equation}
where we used Eq. \ref{Eq13}, and the collapse of the  resonances  into the degenerate energies $E_{(2)}$and $E_{(A)}$. 

The final message is surprisingly intuitive. Molecular dissociation with the liberation of the \(A\) atom occurs at a distance when the coupling between the \(B\) orbital and the substrate exceeds the critical surface bonding \(V_0^{\mathrm{crit.}}\simeq \sqrt{2} V_{AB}\).  
A factor $2$ might have resulted natural in a WBA, because two electrons benefit from the molecular bonding while only one profits from the surface bond. Nonetheless, we obtained $\sqrt{2}$ because we included the energy shift, \(\mathrm{Re}\left[\Sigma_d(\varepsilon)\right]\) , created by the \(d\)-band to infinite order as the principal value of a divergent self-energy. However simple, the justification of this result involved the exact solution of the Dyson equation for the self-energy. This simple result shows that the \textit{molecular dissociation} occurs when the overlap coupling of $B$ to the surface orbital exceeds $\sqrt{2}$  the molecule bonding energy. 

At this critical strength, a bifurcation of the widths of the resonances appears. Beyond that the resonant energy $\widetilde{E}_{b(A2)}\rightarrow E_A $  progressively diminish its width \begin{equation}
\Gamma_{b(A2)}\simeq\left\vert \dfrac{V_{AB}}{V_{0}}\right\vert ^{2}
V_d\underset{V_{0}\rightarrow\infty}{\longrightarrow}0,
\end{equation}
while  $\widetilde{E}_{a(A2)}\rightarrow E_{2}$ and $\Gamma_{a(A2)}\rightarrow 2\Gamma_{d}=2V_d$. 

Indeed, the bonding of the molecule $AB$ to the surface is a smooth continuous process as a function of the surface coupling parameter $V_{0}$. Here, the bonding and antibonding molecular orbitals, $b(BA)$ and $a(BA)$ are continuously transformed into the bonding between the $d_{z^{2}}\equiv$ $1$ surface orbital and the $B$ orbital, $b(1B)$ and $a(1B)$. However, the dissociation of the atom $A$ occurs abruptly at certain critical strength $V_{0}^{\mathrm{crit.}}.$ This is identified with the collapse of two resonances inside the $d$-band (roots of the determinant with imaginary component). These resonances are identified with the bonding and antibonding molecular orbitals $b(A2)$ and $a(A2)$. The real parts of the poles also have a non-analytic dependence on $V_{0}$. The splitting between $b(A2)$ and $a(A2)$ results from the constructive interference between identical through-bond couplings, as described in Ref. \cite{LePaD90}. This, indirect interaction can be seen as a nested Fermi Golden Rule (a state, $A$, that decays into an unstable state $2$, which decays at a rate \(\Gamma_{d}\simeq  V_{{d}}\) (d-band uncertainty) into the \(d \) band). Hence, $A$ becomes more isolated for a stronger $V_{0}$ since the through-bond coupling is $V_{A2}  \simeq  V_{AB}\dfrac{1}{V_{0}}V_{d}$.

Hence, the collapse of the real part of both resonances means also that the coupling $V_{AB}$ becomes irrelevant. Thus, by focusing on atom $A$ we can define the critical coupling $V_{0}^{\mathrm{crit}}$ as the value at which the poles corresponding to $b(A2)$ and $a(A2)$ collapse. Beyond this value they become two degenerate resonances: a narrow one, $\Gamma_{A2}\rightarrow0$, and a broad one that soon becomes $\Gamma_{A2}\rightarrow 2\Gamma_{d}$. The first limit indicates that the orbital $A$ became free from any bond, while the second one manifests that the collective orbital $2$ is no longer part of a resonance but merges into the \(d\)-band. 

\subsection{Molecular dissociation beyond perfect symmetry.}

In the Appendix \ref{A:Asymmetric}, we show that the overall behavior described above is maintained for non-dimer molecules, where the energies are $E_A\gtrsim E_B$ as long as  $|E_A - E_B| \ll V_{AB}$. The original definition of molecular dissociation at the exceptional point where the spectral bifurcation occurs is outside the reachable parameter range. However, that definition could become generalized as the value where the poles of the retarded Green’s function have a \textit{minimal distance}. This new definition includes the exceptional point and is valid for both dimeric and non-dimeric molecules. Thus, molecular dissociation of a non-dimer where \(E_B\approx E_d\) would occur when
\begin{equation}
    V^c_0 \approx \sqrt{2}V_{AB} (1-\tfrac{(E_A-E_B)^2}{8V_d^2}).
\end{equation}
From this expression we learn that in a non-dimer molecule the atom A, whose energy differs more from the \(d\)-band center, is released a bit earlier than in a dimer which is in resonance with the \(d\)-band center. Other parametric situations could be worked out along the same lines.

The fact that the dissociation induced by the catalyst, considered as an environment, is reflected on the non-analytic character of the molecular spectrum suggests the need to design strategies which could test this spectrum. A previous paper \cite{RuDSaPa15} considered the numerical solution for an environment whose single-particle spectrum had the full non-linearity of an actual \(d\)-band together with the complexity of the 2+1+B+A subsystem. However, in the above discussion, by  simplifying the subsystem using a canonical transformation \cite{Ziman69}, we were able to obtain a QDPT just from an analytical solution of a quadratic eigenvalue equation. The critical coupling resulting from this approximation differs in small numerical factors from the  QDPT that would result from the simpler WBA, i.e. the Fermi Golden Rule and the corresponding Lorentzian spectra. This justifies to test scattering properties \cite{Fe58, Moi11, RoBi15} related to a tunneling spectroscopy experiment that occurs under the action of weakly coupled tunneling probes. In an attempt to generalize these results we will  consider the numerical solutions of tunneling current under different environments. The most natural are those with a Lorentzian spectral density, typical from a WBA and Markovian environments. However, we also seek to consider Gaussian environments, as they may account for non-Markovian processes \cite{YaZu21}. Both situations will be considered after discussing the generalities of decoherent tunneling.

\section{Probing Molecular Dissociation through tunneling.}

\subsection{Partially coherent tunneling in a Markovian environment: the D'Amato-Pastawski model.}

The effects of independent Lorentzian environments on each one of the orbitals of a molecule can be studied using the D'Amato-Pastawski model (DPM) \cite{DAPa90,CaFe+Pa14}. Conceptually, the DPM is build as a tight-binding representation of  the Büttiker's idea that a voltmeters induce a decoherence processes. Thus, N local decoherence processes can be represented with the same number of fictitious voltage probes. Within a tight-binding representation these induce N imaginary parts in the site energies that produce  Lorentzian energy broadenings but do not alter much the energy spectrum. However, one has to include the additional condition that the conservation of charge impose N local chemical potentials\cite{Pa91}. In the more formal context, the DPM constitutes a linearization of the self-consistent Keldysh integral equation of the quantum fields\cite{Pa92}. However, even in the linear response, tunneling in presence of decoherence involves a self-consistent evaluation of the current. Being a classical observable, the current dynamics is related to the frequencies, not to the energies. In an open system, the frequencies are no longer uniquely determined by differences between eigenenergies but by the poles of the Keldysh kernel in the frequency domain or, equivalently, the eigenvalues of a Lindbladian\cite{Pa07}. These quantities indeed may show a QDPT \cite{Pa07}. In consequence, it is not clear if a partially coherent tunneling, which accounts for such self-consistency, would present a QDPT. We will devote the next sections to clarify this point.

The local dephasing environment will be represented as an imaginary part $-\rm{i}\Gamma_\phi$ in the molecule's site energies. When this only occurs on one orbital, we would call it \textit{asymmetric} configuration, as it distinguishes left from right side of the molecule. The case where both sites of the molecule have identical independent environments will be addressed as a \textit{symmetric} configuration. In the Keldysh formalism \cite{Pa92}, these imaginary parts can be thought as additional leads whose chemical-potential are set to ensure that no net current flows into them.

It should be noticed that every site could be affected by more than one process. For example, when we are looking at the transference through an atom of the molecule (let's say atom $A$) with a symmetric environment, the site $A$ is affected by three decoherent processes. One corresponding to the actual environment, and the others two accounting for the tunneling electrodes. The sub-indices $S$ and $D$ denotes the source and drain leads, independently if they are in the same site or different sites.

Particularly, the Fisher-Lee formula gives the transmission probability from the process $\alpha$ at position $i$ and the process $\beta$ at site $j$ ($\alpha$ and $\beta$ could be environments/voltmeters or the source and drain leads). In a tight-binding representation\cite{PaMe01,PaFoMe02}:
\begin{equation}
    T_{\alpha i,\beta j}=4\Gamma_{\beta j}G_{j,i}(\varepsilon)\Gamma_{\alpha i}G^*_{i,j}(\varepsilon),
\end{equation}
where $\Gamma_{\alpha i}=\rm{i}(\Sigma_{\alpha i}-\Sigma^*_{\alpha i})/2$ is a magnitude proportional to the escape rate at site $i$ due to a process $\alpha$.

Using the generalized Landauer-Büttiker equations\cite{Pa91}, which describe the balance of electronic current as a Kirchhoff law in terms of the transmittances between different pairs of leads, $T_{\alpha i,\beta j}$, one obtains:
\begin{equation}
    I_{\alpha i}=\frac{e}{h}\sum_{\beta}\sum_{j (\alpha i \neq \beta j)}(T_{\alpha i,\beta j}\delta\mu_{\beta j}-T_{\beta j,\alpha i}),
\end{equation}
where the quantities $\delta\mu_{\alpha i}=\mu_{\alpha i}-\varepsilon_F$, are the changes in the chemical potential of the electron reservoirs at site $i$ when the external electrodes induce non-equilibrium processes.

The condition of no net current in the ``leads'' associated with decoherent process comes in the form $I_{\phi i}=0$, which determines the internal chemical potentials $\delta\mu_{\phi i}$ just by inverting a $M\times M$ matrix, where $M$ is the number of decoherent processes. Accordingly, in the linear response the tunneling current from the source to drain becomes $I_{SD}=\frac{e}{h}\widetilde{T}(\delta\mu_{S}-\delta\mu_{D})$ where:
\begin{equation}
\widetilde{T}=T_{S,D}+\sum_{j,i}T_{S,i}\mathbb{W}^{-1}_{i,j}T_{j,D},
\end{equation}
and the matrix $\mathbb{W}$,
\begin{equation}
    \mathbb{W}_{i,j}=-T_{i,j}+(\sum_{j=S,D,\phi_j}T_{i,j})\delta_{i,j}.
\end{equation}

It is clear that the effective transmittance $\widetilde{T}$ is the sum of a coherent transmittance $T^{\mathrm{coh.}}_{S,D}=T_{S,D}$ and a incoherent transmittance $T^\mathrm{inc.}_{S,D}=\sum_{j,i}T_{S,i}\mathbb{W}^{-1}_{i,j}T_{j,D}$ that represents the current of particles that suffered the interaction with the environments.

In the subsections that follow we will study the current as a function of the Fermi energy \(\varepsilon\) of the perturbative electrodes. This implies that one has a plunger-gate electrode that allows the measurement of \(\widetilde{T}(\varepsilon)\). Indeed, this would involve other non-trivial changes, such as Coulomb blockade and effects that lead the system beyond the simplistic Hartree-Fock description we adopted so far\cite{DALP07}. However, we will disregard those changes for the sake of simplifying our exploration.

\subsubsection{Through-atom tunneling with an asymmetric environment.}

In the particular case of a dimer with an environment at site $B$ (asymmetric environment) and probing leads at site A, the transmittance in the DPM is calculated as follows. We evaluate the Green's Functions $G_{AA}$ and $G_{AB}$,

\begin{eqnarray}
G_{AA}(\varepsilon)&=&\frac{1}{\varepsilon-E_0-\dfrac{V_{AB}^2}{\varepsilon-E_i}-2 \Sigma_l}\\
&=&\frac{1}{\varepsilon-E_0-\dfrac{V_{AB}^2}{\varepsilon-E_B+\rm{i}\Gamma_\phi}-2\Sigma_l},
\end{eqnarray}

where the environment at $B$ is introduced by a complex number $E_B-\rm{i}\Gamma_\phi$, and $2\Sigma_l$ accounts for the self-energies of the \textit{leads}. Eventually this could be written in terms of the energy center of the lead \(E_l\), the hopping inside the lead \(V\) and the hopping \(V_l\) which describes the tunneling into it. Similarly, for $G_{AB}$, we have:
\begin{equation}
G_{AB}(\varepsilon)=\frac{V_{AB}}{(\varepsilon-E_0-\Sigma_l)(\varepsilon-E_B+\rm{i}\Gamma_\phi-\Sigma_{l})-V_{AB}^2}.
\end{equation}
In both cases $E_A=E_B=E_l=0$ and thus $\Sigma_{l}=\frac{V_l^2}{V^2}\Sigma(0)=-\rm{i}\Gamma_{l}$ (wide band approximation). 

Therefore, the effective transmittance (i.e. the sum of the coherent and incoherent transmittances) is:
\begin{eqnarray}
\widetilde{T}&=&2\Gamma_{A1}|G_{AA}|^2 2\Gamma_{A2}+2\Gamma_{A1}\frac{2\Gamma_{B\phi}|G_{AB}|^2}{2(2\Gamma_{A1}+2\Gamma_{A2})}2\Gamma_{A2}\\
\widetilde{T}&=&4\Gamma_{l}(\Gamma_{l}|G_{AA}|^2+\Gamma_{\phi}\frac{|G_{AB}|^2}{2}).
\end{eqnarray}
From this we see that analytic properties of the transport observables are already encoded the modulus square of the different retarded Green's function matrix elements.

\subsubsection{Through-atom  tunneling with a symmetric environment.}

When there are environments in both sites of the molecule (symmetric environment), the effective transmittance becomes:

\begin{equation}
    \widetilde{T}=T_{A_S,A_D}+\\
    \begin{pmatrix}
T_{A_S,\phi_1} \\
T_{A_D,\phi_2}  
 \end{pmatrix}^{T}
\mathbb{W}^{-1}
\begin{pmatrix}
T_{A_S,\phi_1} \\
T_{A_S,\phi_2}  
 \end{pmatrix}
\end{equation}  
where
\begin{equation}
    \mathbb{W}=\begin{pmatrix}
T_{A_S,\phi_1}+T_{A_D,\phi_1}+T_{\phi_1,\phi_2} & T_{\phi_1,\phi_2} \\
T_{\phi_1,\phi_2} & T_{A_S,\phi_2}+T_{A_D,\phi_2}+T_{\phi_1,\phi_2}
\end{pmatrix}
\end{equation}
To simplify the notation, we can notice that  $T_{A_S,\phi_k}=T_{A_D,\phi_k}=T_{A,\phi_k}$, for $k=\{1,2\}$ and defining:
\begin{eqnarray}
W&=&(2T_{l,\phi_1}+T_{\phi_2\phi_1})(2T_{l,\phi_2}+T_{\phi_1\phi_2})-(T_{\phi_1\phi_2})^2\\
\left[\mathbb{W}^{-1}\right]_{1,1}&=&(2T_{l,\phi_2}+T_{\phi_2\phi_1})/W\\
\left[\mathbb{W}^{-1}\right]_{2,2}&=&(2T_{l,\phi_1}+T_{\phi_1\phi_2})/W\\
\left[\mathbb{W}^{-1}\right]_{1,2}&=&(T_{\phi_1\phi_2})/W\\
\end{eqnarray}

\begin{eqnarray}
\widetilde{T}&=&T_{A_S,A_D}+\begin{pmatrix}
T_{l,\phi_1} & T_{l,\phi_2}  \\
 \end{pmatrix}
\begin{pmatrix}
\left[\mathbb{W}^{-1}\right]_{1,1} & \left[\mathbb{W}^{-1}\right]_{1,2} \\
\left[\mathbb{W}^{-1}\right]_{1,2} & \left[\mathbb{W}^{-1}\right]_{2,2}
\end{pmatrix}
\begin{pmatrix}
T_{l,\phi_1} \\
T_{l,\phi_2} 
 \end{pmatrix}\\
 &=&T_{A_S,A_D}+\frac{T_{A,\phi_1}+T_{A,\phi_2}}{2}\\
 &=&4\Gamma_l^2|G_{A,A}(\varepsilon)|^2+2\Gamma_l\Gamma_\phi(|G_{A,A}(\varepsilon)|^2+|G_{A,B}(\varepsilon)|^2)
\end{eqnarray}
In this case, as the environment induced broadening $\Gamma_\phi$ becomes stronger, the peaks in the transmittance also become wider until it is impossible to distinguish one from the other. In contrast to the case of the asymmetric environment, this ``single peak'' gets spread over a greater range of energies, and therefore no-critical behavior or phase transition can be trivially associated to this observable. 

\subsection{Beyond the Markovian environment: the Gaussian  broadening.}

In the description above, all the analytical calculations were based on the simple mathematical properties of the Lorentzian line broadening that result from the wide band approximation or the Fermi Golden Rule. They give the same conceptual picture as the exact numerical results obtained in Ref. \cite{RuDSaPa15,RuDSaPa16}. Those solutions involved the fully non-linear equations and complex forms of the uncertainties, as the Lorentzian tails are naturally truncated by the band edges of the metallic substrate\cite{DeBMPa08}. In that case, the evaluation of the poles of the Green's functions and resonant states could be evaluated from the analytic structure of the Dyson equation.

Thus, one natural question is whether the QDPT persists when the interactions with the environment has non-Markovian memory effects. In such cases, the self-consistent solution of the Keldysh equations become almost inaccessible. One could benefit neither from the analytical structure of the Green's functions nor from the simplicity of the Fermi Golden Rule or the simple analytical dependence of the self-energies. Therefore, in the rest of the paper, we want to introduce a solvable model that could bypass these difficulties. It has two steps: 

\begin{enumerate}
\item A particular environment model that could provide the Gaussian uncertainties. Such uncertainty involves enough short time memory as to lead to non-Markovian effects. This is the actual situation often observed in experiments with a spin bath \cite{ADLP06,Sa+Pa20}. The idea is that such bath could result from the limit of the binomial distribution, which in turn can be represented as a binary Galton's board. That would result from the sum of binary energies when an orbital couples with a large set of fictitious spins or qubits that take the value 0 and 1. That sort of model is discussed in the time domain in Refs.  \cite{ZuCP07,YaZu21}. 
\item A numerical method that could be used to test the molecular spectrum of a molecule interacting with different environments. This model should be tested against environments that yield Lorentzian uncertainties in the molecular states. Here, the basic idea will be to study the resonances observed in the perturbative current that results when the molecule, dressed by its environment, is placed under the electrodes of an ideal tunneling microscope.
\end{enumerate}

In this direction, our proposal for the environment is:
\begin{align*}
\hat{H}_{\mathcal{E}} &  =\sum_{k}\Omega_{k}\hat{S}_{k}^{+}\hat{S}_{k}^{-},\\
\hat{V}_{B\mathcal{E}} &  =\sum_{k=1}^{N}V_{s}\hat{c}_{B}^{+}\hat{c}%
_{B}\left[  \hat{S}_{k}^{+}+\hat{S}_{k}^{-}\right] ,\\
V_{s} &  =\sigma_{\phi}/\sqrt{N}.\\
\end{align*}

The last election ensures that, in our model, the second moment of the perturbation remains constant regardless of the number of spins that we adopt. In order to visualize the consequences of our model we may represent the case of two orbitals and two spins in three different configurations (Fig. \ref{System}): 
\begin{itemize}
\item \textbf{Asymmetric} environment, where only \textbf{one} \textbf{orbital}  is in contact with the environment.
\item \textbf{Symmetric} environment, were \textbf{each orbital}  is in contact with a different environment.
\item \textbf{Correlated environment}. Where \textbf{both orbitals} can interact with unique set of spins. It will not be considered here for brevity.
\end{itemize}

The interesting property that facilitates the calculation is to exploit once more the \textit{conservation of the orbital symmetry}  \cite{WoHo69}. This ensures that there is a common basis that diagonalizes the spin interactions within each \textit{layer of the Fock-space}\cite{PaFoMe02} associated with a given orbital (\textit{e.g.} $ \left\vert A00\right\rangle $, $ \left\vert A01\right\rangle $, $ \left\vert A10\right\rangle $ and $ \left\vert A11\right\rangle $) then the bonding interaction \textit{only mixes} those collective states with the \textit{same symmetry} at different layers. A compilation of the notation, magnitudes, and physical meanings is displayed at Table 1.

\begin{table}
\begin{center}
\begin{tabular}{||c || c||} 
\hline
$V_{AB}$ &  \text{molecular bonding energy per electron}\\\hline
\ $V_d\gg V_{AB}$ & d\text{-banding effective interaction (Lanczos basis)}\\\hline
 $\Sigma_{d}$ & self-energy of the surface $d_{z^{2}}$ orbital \\ &  (environment that enables catalysis)\\\hline
$\Gamma_{\phi}$&\text{Lorentzian uncertainty/broadening}\\\hline
$\sigma_{\phi}$& \text{Gaussian uncertainty/broadening}\\\hline
$V_{s} =\sigma_{\phi}/\sqrt{N}$& \text{interaction with each \textquotedblleft
spin\textquotedblright} $V_{s}\hat{c}^{+}\hat{c}\left(  \hat{S}_{j}^{+}+\hat{S}_{j}^{-}\right)$  \\\hline
$\hat{H}_{\mathcal{E}}={\displaystyle\sum_{j=1}^{N}}
~\Omega_{j}\hat{S}_{j}^{+}\hat{S}_{j}^{-}$ & \text{spin bath}\text{with Zeeman energy }$\Omega_{j}\equiv0$\\\hline
$\hat{V}_{\mathcal{SE}}={\displaystyle\sum_{j=1}^{N}}V_{s}\hat{c}_{A}^{+}\hat{c}_{A}\left(\hat{S}_{j}^{+}+\hat{S}_{j}^{-}\right)$&   \text{ perturbation
operator}\\\hline
$E_{i}=\left(N-2i\right)\sigma_{\phi}/\sqrt{N}$  &   \text{\ \  Fock's space local energies
}\\\hline
$\eta$ &  \text{broadening/regularization of divergences at eigenenergies }\\\hline
$\Sigma_{l}=\left\vert V_{l}/V\right\vert ^{2}\Sigma_{{}}$ & \text{electrodes tunneling
self-energy}\\\hline
$=\Delta_{l}-\mathrm{i}\Gamma_{l}$& \\\hline
$\Delta_{l} =0$\text{ and  }$\Gamma_{l}=$\text{const.}&\text{wide band approximation}\\
\hline
\end{tabular}
\end{center}
\caption{Here, we summarize the main parameters used in the analytical and numerical calculation making explicit their physical meanings.}
\end{table}

\subsubsection{Transmittance through the Fock-space (orbitals+spin bath) .}

\begin{figure}
\centering
\includegraphics[width=0.9\textwidth]{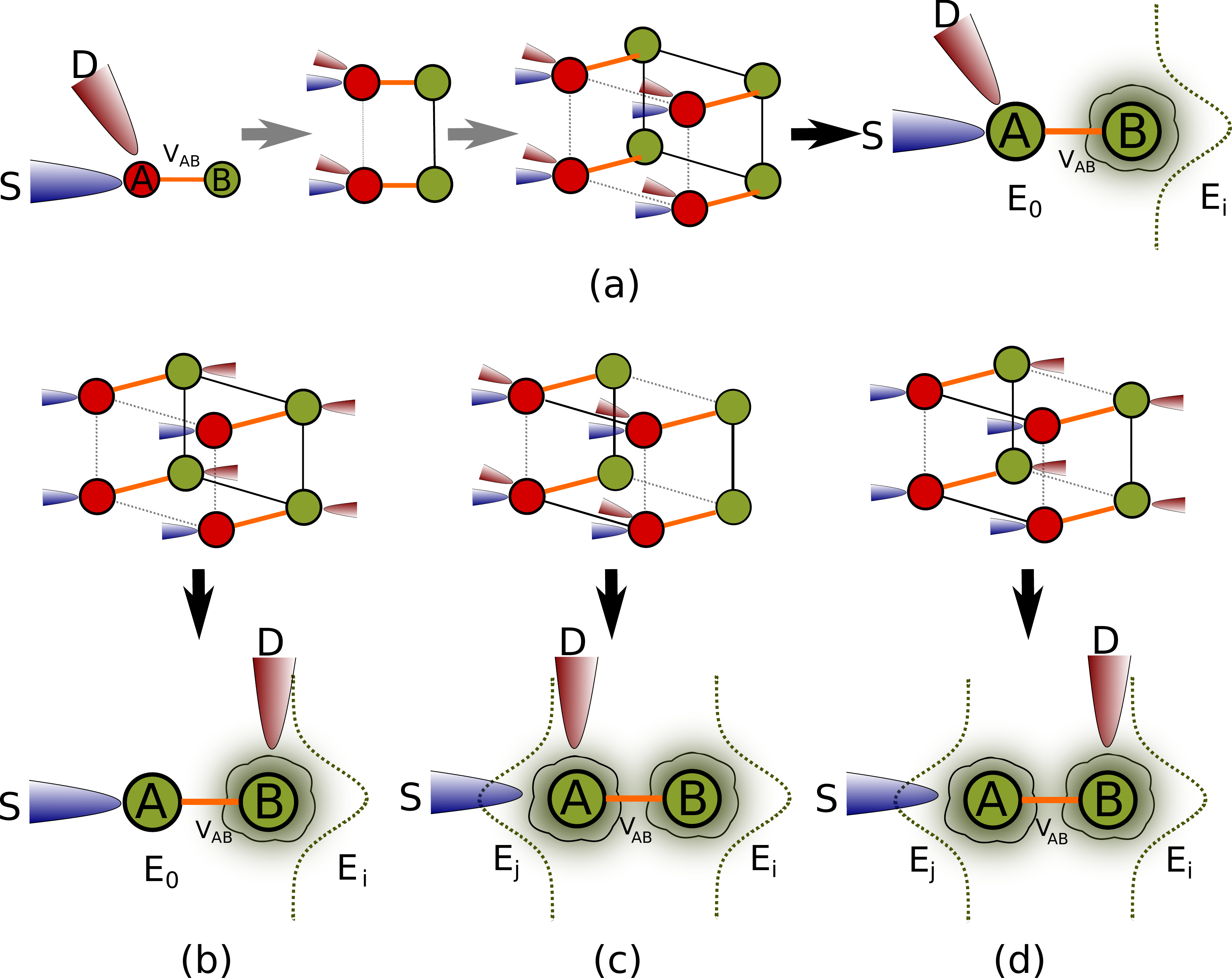}
\caption{Transmittance through the Fock space. \textbf{(a) }A two-site molecule ($A$ (Red) and $B$ (Green)) with a source and a drain lead connected at site $A$ (transport through-atom). In this case, only site ($B$) is affected by a spin environment (asymmetric environment). From left to right the environment generates more configurations (Fock states) at site $B$ as the number of spins increases. An equivalent representation for large environment is represented by the Gaussian broadening at the extreme right. Figures a-c represent different transport configurations in the molecule, the top figure is a representation in the Fock space while the bottom figures an equivalent representation of the model. \textbf{(b)} Through-bond transmittance  with an asymmetric environment. \textbf{(c)} Through-atom and  \textbf{(d)} through-bond transmittances both with a symmetric environment.}
\label{System}
\end{figure}

Let's consider the system in figure \ref{System}-(a), which represent a molecule, with a spin environment at site $B$, connected to two leads (Source and Drain) at site $A$. The Hamiltonian of the system will be $\mathbb{H}=\begin{pmatrix}
\mathbb{H}_A & -V_{AB}\mathbb{I} \\
-V_{AB}\mathbb{I} & \mathbb{H}_B
\end{pmatrix}$, with $\mathbb{H}_A=\Sigma(\varepsilon)\mathbb{I}$ a diagonal Hamiltonian, with the self-energies of the leads at its diagonal and $\mathbb{H}_B$ a Hamiltonian that contains the hopping between the states of the bath (hyper-cubes of $2^N$ sites represented as Green sites in Fig. \ref{System}-a)). In this basis, the transmittance, is given by:
\begin{eqnarray}
T(\varepsilon)&=&4\rm{Tr}[\bbGamma\mathbb{G}\bbGamma\mathbb{G}^{\dagger}]=\sum^{2^{(N+1)}}_{\alpha=1}\langle\alpha|\bbGamma\mathbb{G}\bbGamma\mathbb{G}^{\dagger}|\alpha\rangle\\
&=&4\Gamma\sum^{2^N}_{\alpha=1}\langle\alpha|\mathbb{G}\sum^{2^{N+1}}_{\beta=1}|\beta\rangle\langle\beta|\bbGamma\mathbb{G}^{\dagger}|\alpha\rangle\\
&=&4\Gamma^2\sum^{2^N}_{\beta=1}\sum^{2^N}_{\alpha=1}\langle\alpha|\mathbb{G}|\beta\rangle\langle\beta|\mathbb{G}^{\dagger}|\alpha\rangle=4\Gamma^2\sum^{2^N}_{\beta=1}\sum^{2^N}_{\alpha=1}|G_{\alpha,\beta}|^2\\
\\&=&4\Gamma^2(\varepsilon)2^N\sum_{\beta}|G_{0,\beta}|^2,
\end{eqnarray}
where $G_{\beta,\alpha}=\langle \beta|\mathbb{G}|\alpha\rangle=\langle \beta|\frac{1}{\varepsilon-\mathcal{H}}|\alpha\rangle$, and $\alpha$ and $\beta$ run over Fock states in the molecule $A$ (represented as red sites in Fig. \ref{System}). We also used that $\bbGamma_{i,j}=0$ if $i\neq j || i=j>2^N$ and $\bbGamma_{i,i}=\Gamma(\varepsilon)$ for $i\leq2^N$. As it is stated, this problem involves the inverse of a $2^{N+1}$ matrix. However, this can be simplified by using the appropriate basis. Consider the matrix $\mathbb{U}_{p}$ that diagonalizes $\mathbb{H}_B$, i.e. $\mathbb{H}^D_B=\mathbb{U}_p \mathbb{H}_B \mathbb{U}_p^{-1}$ is diagonal, and extend it to the following matrix: $\mathbb{U}=\begin{pmatrix}
\mathbb{U}_p & \mymathbb{0} \\
\mymathbb{0} & \mathbb{U}_p
\end{pmatrix}$. Then, $\mathbb{U}\mathbb{H}\mathbb{U}^{-1}=\begin{pmatrix}
\mathbb{H}_A & -V_{AB}\mathbb{I} \\
-V_{AB}\mathbb{I} & \mathbb{H}^D_B
\end{pmatrix}$, which have transformed the Hamiltonian in a set of $2\times2$ Hamiltonians. Therefore, we can express the Green function and the transmittance as follows:
\begin{eqnarray}
\bar{\mathbb{G}}=\mathbb{U}^{-1}\mathbb{G}\mathbb{U}=\frac{1}{\varepsilon\mathbb{I}-\mathbb{U}\mathbb{H}\mathbb{U}^{-1}},
\end{eqnarray}

\begin{eqnarray}
T(\varepsilon)&=&4\rm{Tr}[\bbGamma\mathbb{U}\mathbb{U}^{-1}\mathbb{G}\mathbb{U}\mathbb{U}^{-1}\bbGamma\mathbb{U}\mathbb{U}^{-1}\mathbb{G}^{\dagger}\mathbb{U}\mathbb{U}^{-1}]\\
&=&4\rm{Tr}[\mathbb{U}^{-1}\bbGamma\mathbb{U}\bar{\mathbb{G}}\mathbb{U}^{-1}\bbGamma\mathbb{U}\bar{\mathbb{G}}^{\dagger}]\\
&=&4\rm{Tr}[\bbGamma\bar{\mathbb{G}}\bbGamma\bar{\mathbb{G}}^{\dagger}]\\
&=&4\Gamma^2\sum^{2^N}_j\sum^{2^N}_{i}|\bar{G}_{i,j}|^2=4\Gamma^2\sum^{2^N}_{i}|\bar{G}_{i,i}|^2,
\end{eqnarray}
where we used that $\bbGamma$ do not change with the transformation and that, as $\mathbb{H}$ in this new basis is a set of $2\times 2$ Hamiltonians, $G_{i,j}=0$ for $i\neq j$ while $i\leq 2^N \& j\leq 2^N$.

From the expression of the transmittances in the two basis follows:
\begin{eqnarray}
T_0(\varepsilon)&=&4\Gamma^2\sum_{\alpha}|G_{0,\alpha}|^2\\
&=&4\Gamma^2\frac{1}{2^N}\sum^{2^N}_{i}|\bar{G}_{i,i}|^2.
\end{eqnarray}
To round up, we have expressed the transmittance as a sum of two sites transmittance. In this case (Fig. \ref{System}-a), the ``atom'' $A$ remains with the same site energy, while for ``atom'' $B$ the energies $E_i$ are given by:
\begin{equation}
    E_i=(N-2i)\times \sigma_{\phi}/\sqrt{N},
\end{equation} where the denominator is a consequence of the form of the interactions and normalizes the width of the energy distribution. The degeneracy of these energies is given by the Binomial coefficients, $W_i=\binom{N}{i}$, which lead to a Gaussian distribution of $E_i$ as $N$ increases.

To account for the transport through this system, we  add to the total transmittance $\widetilde{T}$, the transmittances for each energy $E_i$ weighted by their degeneration:

\begin{equation}
    \widetilde{T} = \sum_i W_i\times T_i,
\end{equation}where, 
\begin{equation}
  T_i = \frac{4}{2^N}\Gamma_L |G_{LR}(\varepsilon)|^2 \Gamma_R.
\end{equation}

When the tunneling current occurs in a \textbf{through-atom} configuration (Fig. \ref{System}-a and \ref{System}-c), the Green's function corresponds to the Green's function of site $A$ after the decimation of the environment $i$,

\begin{equation}
G_{A_iA_i}(\varepsilon)=\frac{1}{\varepsilon-E_0-\frac{V_{AB}^2}{\varepsilon-E_i}-2\Sigma_l},
\end{equation}where  $\Sigma_l$ is the self energy of the leads in the wide band approximation.

On the other hand, when the transport is \textbf{through-bond} (Fig. \ref{System}-b and \ref{System}-d) the Green's function becomes:
\begin{equation}
G_{AB}(\varepsilon)=\frac{V_{AB}}{(\varepsilon-E_0-\Sigma_l)(\varepsilon-E_i-\Sigma_l)-V_{AB}^2}.
\end{equation}
When the environment is symmetric (\ref{System}-c and \ref{System}-d), the generalization is straight-forward by $E_0\rightarrow E_j$ and summing over $j$ with an extra factor $W_j/2^N$.

\subsection{Numerical results in different tunneling configurations.}

In this section, we numerically analyze the through-bond and through-atom transmittance for asymmetric and symmetric environments. The results will be focused on the behavior of $T$ as the environments strength $\Gamma_\phi$ and $\sigma_\phi$ increase. The leads are very weakly connected to the atom ($V_l=0.05V$) which results in $\Gamma_l=0.0025V$, i.e. deep into the wide band limit. The molecule bond is fixed at $V_{AB}=0.1V$. From here on, the environment strength and energies will be referred to $V_{AB}$.

\subsubsection{Asymmetric environment.}
  
As it was discussed in the previous sections, an asymmetric Lorentzian environment produces a QDPT breaking the bond of the molecule at a critical environment strength. For a Gaussian environment, generated from the bath described above, some differences appear.
 
\begin{figure}
\centering
\includegraphics[width=0.9\textwidth]{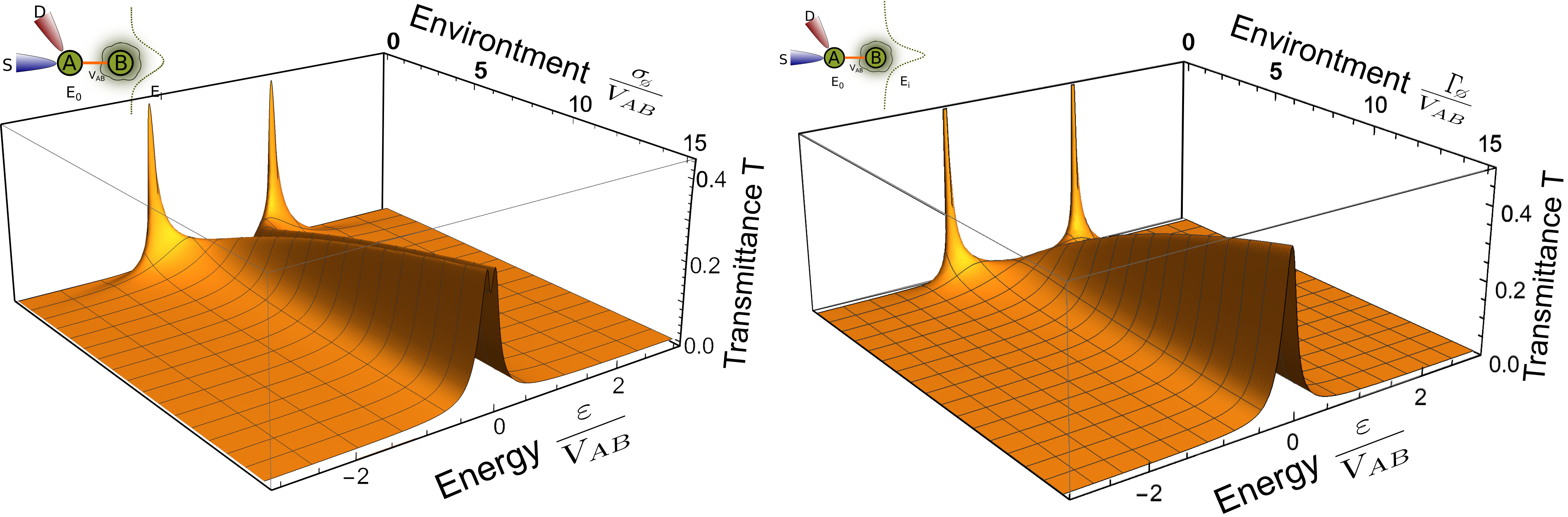}
\caption{Through-atom tunneling. Transmittance $T$ as a function of the energy and the  (a) Gaussian environment strength $\sigma_{\phi}/V_{AB}$. (b) Lorentzian environment strength $\Gamma_{\phi}/V_{AB}$. }
\label{Fa-1P-Q}
\end{figure}

As the environment strength $\sigma_\phi$ increases from 0 to 1.5 in $V_{AB}$ units (Fig. \ref{Fa-1P-Q}-a) we can observe  several regimes. First, the resonances' transmittance decrease and their width increase. For larger $\sigma_\phi$, the resonances move to the center while maintaining the two maxima and increasing the transmittance. In this case, although the width of the peaks decreased, no collapse into one resonance is observed. The decreasing width of the peaks as well as its displacement to the center are signs of an increasing isolation of the states. As well as in the Lorentzian case (Fig. \ref{Fa-1P-Q}-b), the bonding between $A$ and $B$ decreases, however, no critical behavior is evident. We interpret this as a particularity of the Gaussian model adopted that has an antiresonance at \(\varepsilon=0\).

The absence of a critical breaking shows that the coupling $V_{AB}$ between $A$ and $B$ does not become irrelevant, i.e. some small residue of the bonding and antibonding signature still persist. This is observed as a minimum at $\varepsilon=0$, a direct consequence of the particular model  we used to represent a Gaussian environment. The light-tails of the Gaussian distribution generate big energy fluctuations with an exponentially decreasing probability. This is reflected as the appearance of two modes in the statistical distribution of the inverse of the collective eigenenergies. This property is chased down to its effects on the transmittance (for a detailed deduction see \ref{A:GaussEnv}). Mathematically, it might seem that ``heavy-tailed'' strong fluctuations of the Lorentzian environment play some role in the molecule dissociation. However, the occurrence of the phenomenon in a model with a finite bandwidth contradicts this hypothesis. A similar model that samples energies with a Lorentzian distribution was tested numerically and analytically, showing an equivalence with the D'Amato-Pastawski model (See \ref{A:LorEnv}) similarly to what occurs with resonant tunneling model \cite{PaFoMe02}.

\begin{figure}
\centering
\includegraphics[width=0.5\textwidth]{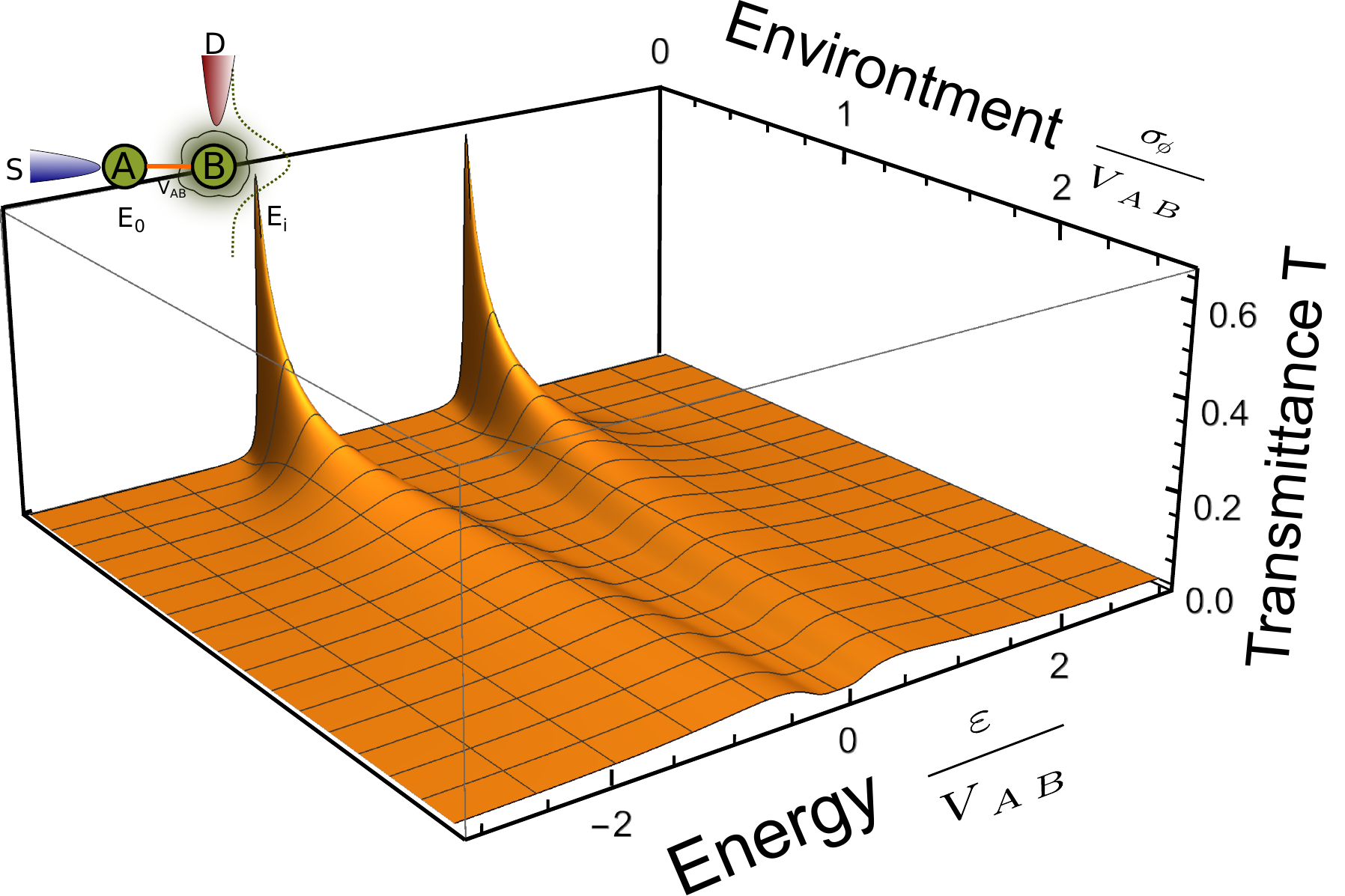}
\caption{\textbf{Through-bond} tunneling with an asymmetric Gaussian environment,  transmittance $T$ as a function of the energy and the environment strength $\sigma_\phi$.}
\label{Fb-vs}
\end{figure}

Now, if we consider that the leads are placed at different atoms of the molecule, i.e. \textbf{through-bond} transport necessarily, the qualitative physics is different. When we vary the environment strength $\sigma_\phi$, in contrast to the preceding example, the transmittance fades away as the transport through through very off-resonant ``collective orbitals''  is hindered. There is always an antiresonance at $\varepsilon=0$ and the transmittance decreases with $\sigma_\phi$ (Fig. \ref{Fb-vs}). If $\sigma_\phi$ is big enough, most of the $E_i$ levels are decoupled from the chain, and only a few of them (the center of the distribution) allow transport. The statistical significance of these states becomes negligible as $\sigma_\phi$ increases. For Lorentzian environments (DPM) the antiresonance at the $\varepsilon=0$ comes from the incoherent transmittance and it is mounted on the coherent transmittance, where a collapse of the peaks is observed. Nevertheless, both the coherent and incoherent transmittance fade away as $\Gamma_\phi$ increases. For a more complete analysis of the symmetric Lorentzian through-bond tunneling see Appendix \ref{A:TBLQ}.

\subsubsection{Symmetric environments.}
 
\begin{figure}
\centering
\includegraphics[width=0.5\textwidth]{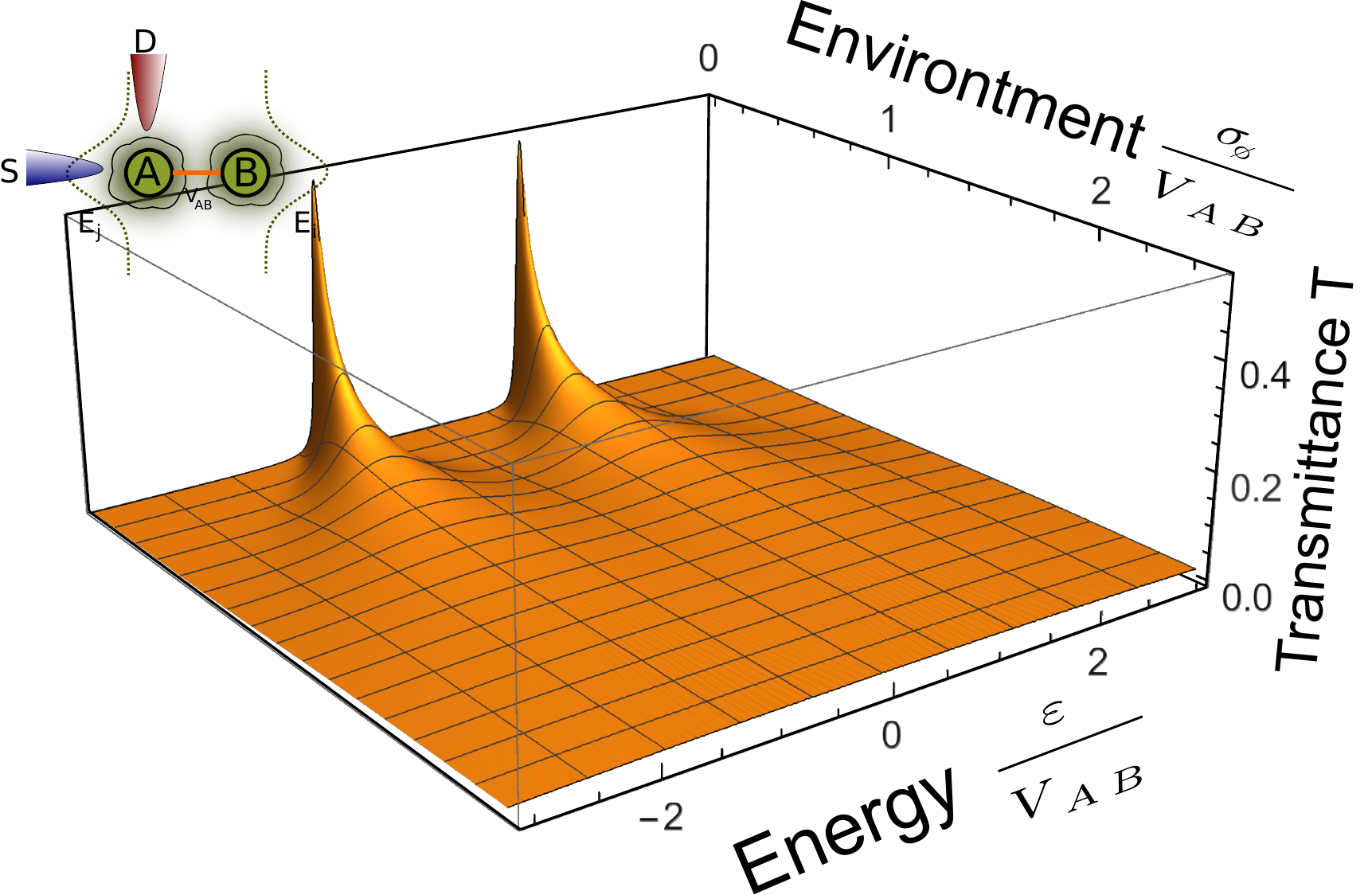}
\caption{\textbf{Through-atom} tunneling with symmetric Gaussian environments. Transmittance $T$ as a function of the energy for different environment strength $\sigma_\phi$.}
\label{Fb-vs-nq-T-a}
\end{figure}

For symmetric environments, the transmittance decreases as their strength increases. As in the asymmetric through-bond case, the number of energy configurations that contribute to the total transmittance becomes negligible. However, in this case, the minimum at the center of the energies eventually disappears. In this case, no real transition can be associated to the unification and can be (simply and empirically) fitted by the sum of two Gaussians:

\begin{equation}
    DG(\varepsilon)=H(e^{-\frac{(\varepsilon-\varepsilon_0)^2}{2\sigma^2}}+e^{-\frac{(\varepsilon+\varepsilon_0)^2}{2\sigma^2}})
    \label{doubleGauss}
\end{equation}

When $\varepsilon_0$ does not depend on $\sigma$,   $DG(\varepsilon)$ has a change in its concavity at $\varepsilon = 0$ for a finite $\sigma$. Therefore, the two maxima at $\varepsilon \neq 0$ become a maximum at zero.  In our model  $\varepsilon_0$ change slightly with $\sigma_\phi$ and therefore with $\sigma$, nevertheless, this change in the concavity still occurs.

Finally, $\varepsilon_0(\sigma_\phi)$ shows how the center of each Gaussian moves away as $\sigma_\phi$ increases, although the maxima of the transmittance move to $\varepsilon=0$. For Lorentzian environments (DPM) similar results are obtained.

\subsection{Final remarks on tunneling under Lorentzian and Gaussian environments.}

\begin{figure}
\centering
\includegraphics[width=0.9\textwidth]{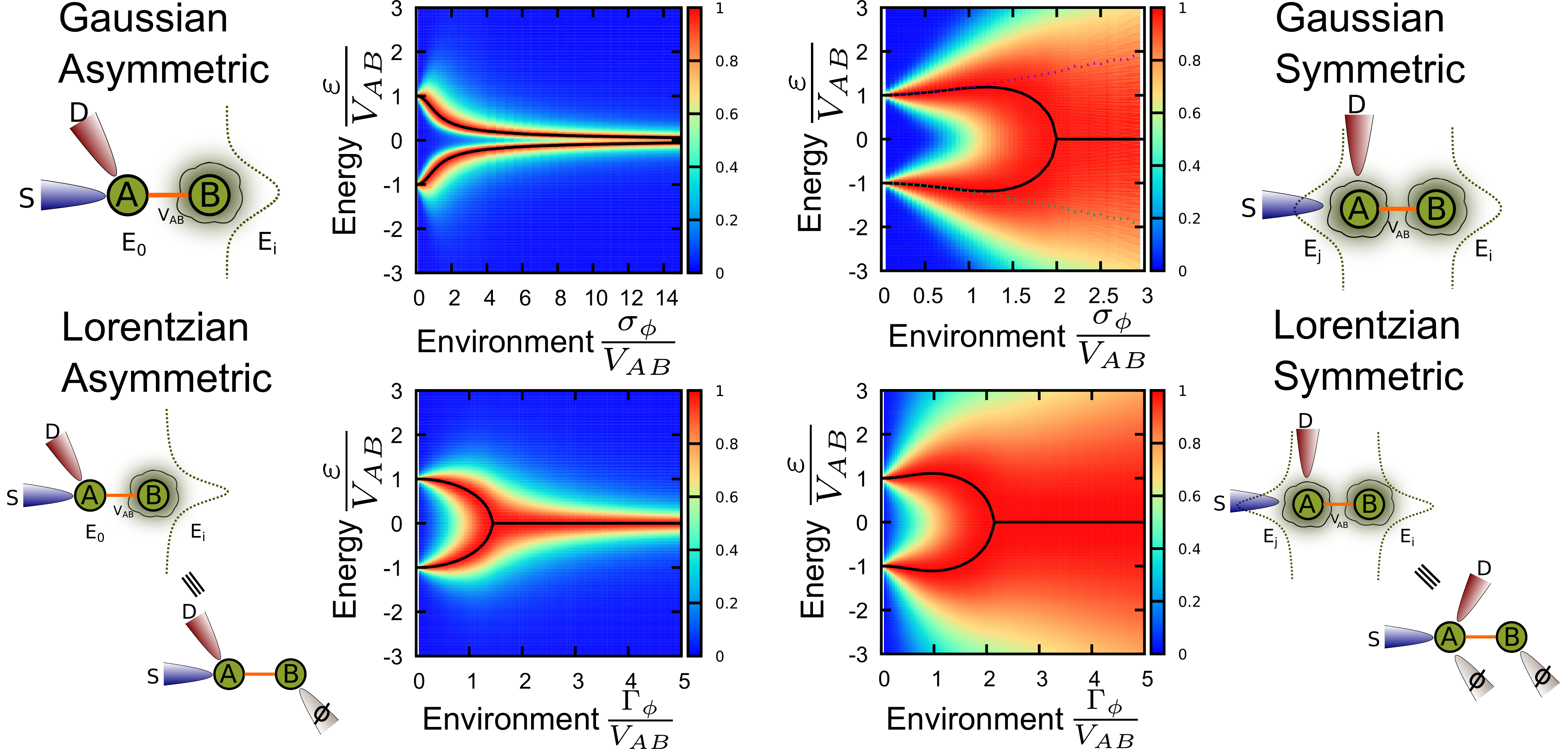}
\caption{Comparison of the normalized transmittance in the four studied cases. Top: Gaussian environment. Bottom: Lorentzian environment. Left and Right: Asymmetric and symmetric environments respectively. The solid black lines show the maximum of the transmittance. In the Gaussian-Symmetric plot the dashed lines display the center of the two Gaussian fit with Eq. \eqref{doubleGauss}.}
\label{Compara}
\end{figure}

To summarize the differences and similarities of the effects of Gaussian and Lorentzian environments on the tunneling transmittance, in Fig. \ref{Compara}, we show the normalized transmittance $T$ as a function of the environment strength in four paradigmatic cases: through-atom transport with symmetric and asymmetric Gaussian and Lorentzian environments. The plots include the maxima of the transmittance as a function of the environment strength as a solid-black line. As it is discussed in Appendix \ref{A:LorEnv}, the Lorentzian environment is equivalent to the D'Amato-Pastawski model. This is represented in the schematic figures beside each plot.

On one hand, by contrasting the normalized \textbf{through-atom} transmittance in the \textbf{asymmetric} case (Left column of Fig. \ref{Compara}), it is clear how, in the \textbf{Lorentzian} case, a collapse of the transmittance is observed for a finite environment strength. This confirms the QDPT. At this point, it is important to notice that the numerical critical strength of the environment results $\Gamma \simeq \sqrt{2}V_{AB}$ instead of the $\Gamma = {2}V_{AB}$ that one may naively expect from the eigenvalues of the non-Hermitian Hamiltonian.  The previous results of the Lorentzian environment also contrast with those of the the \textbf{Gaussian} one. The transmittance peaks approach to zero while they always maintain a two peak behavior. In both cases, as the environment increases its strength, all the transmittance concentrates in the surroundings of the Fermi energy, i.e., there are no actual bonding or antibonding resonances. Also, they have a decreasing line-width indicating that the molecular dissociation, while not critical, indeed occurs.

On the other hand, in the \textbf{symmetric} configuration, where both atoms are coupled to a environment (Right column of Fig. \ref{Compara}), the behavior of the \textbf{through-atom} transmittance under either Lorentzian or Gaussian environments  fades away  and becomes wider as the environment strength increases. In these cases, the change in the concavity (minimum/maximum) at the center is roughly described as the sum of two resonances whose second moment increases. For the Lorentzian case the critical behavior yields a value that coincides with the dynamical analysis of Keldysh or Lindbladian with a critical value for dissociation of  $\Gamma = 2V_{AB}$\cite{FePa15,Pa07}. For the Gaussian environment, the transmittance was fitted with Eq. \eqref{doubleGauss}, the Gaussian centers, $\sim\pm \varepsilon_0$, are shown as dashed-lines over the plot. We observe how the centers of the individual peaks repel each other as the environment strength increases. However, their increasing widening still produce a maximum at the center of the distribution.

\section{Conclusions.}
We first presented a conceptual review on how a dimeric molecule dissociation (formation) in presence of a catalyst can be viewed as a Quantum Dynamical Phase Transition (QDPT). This, in turn, has a mathematical correspondence with the spectral discontinuity (bifurcation/collapse) of the eigenvalues of the effective Hamiltonian as the dimer approaches the catalyst. Indeed, this is a notable situation involving a spontaneous symmetry breaking induced by the coupling with the \(d\)-band of the catalyst that is in precise resonance with the dimer. When the molecule is far away from the catalyst, full symmetry should be reached. Even a slight departure from the molecular symmetry, e.g. a general $AB$ or non-dimer molecule, would smooth out the non-analytic discontinuity. Such regularization of the singularities is equivalent to what occurs with van Hove singularity of the density of states of a solid, i.e., $N(\varepsilon)=-(1/\pi) \lim_{\eta \rightarrow 0}\rm{Im}G(\varepsilon +\rm i \eta)$, that smooths out \(N(\varepsilon)\) in the presence of a small energy uncertainty $\eta$. Thus, more generally, molecular dissociation should be assimilated to a minimal distance between internal resonant states. In any case, the very existence of the spectral exceptional point, allowed us to establish reliable estimates for the parameters for which the molecule dissociation (formation) occurs, as expressed in Eqs. 13-17. These  give specific bounds and generality to the numerical evidence shown in Ref. \cite{RuDSaPa15}.

Having set the parameters of a dimer under action of an \textit{effective asymmetric environment} that could manifest a QDPT, we used them in a numerical test that evaluates the effects of the particular forms of the environment in the transmittance. The idea is that this test could be assimilated to an experimental tunneling microscopy set-up that could be used to test the particular forms of the environment.
Eventually, from this, one could infer or even induce a critical behavior. With this purpose, we studied the transmission spectrum in presence of different environments by analyzing the tunneling through one atom (through-atom) or through the whole effective dimer (through-bond) as a magnitude that could reflect the molecular dissociation. 
For a Lorentzian environment, we obtained results that reproduce those of a full \(d\)-band of Ref. \cite{RuDSaPa15,EGM20}, where it was shown that the environment in one of the atoms produces a QDPT. We compare the energy uncertainty induced by the environment, which has a dynamical interpretation as energy fluctuation\cite{FePa15}, and the D'Amato-Pastawski model (DPM) for steady state transport demonstrating that, in this system, both approaches are equivalent. The difference between the critical values on the transmittance and the Hamiltonian spectrum draws out attention to a frequently ignored feature:  the actual experimental observables should be evaluated from the Keldysh (here DPM with GLBE) or Lindbladian equation and thus their main features would not necessarily coincide with the eigenvalues of the Hamiltonian. In a metaphorical way: \textit{the observables are a particular shadow  of the Hamiltonian spectrum}.
 
We also introduced a new solvable Hamiltonian model that, in the thermodynamic limit, generates the energy broadening that results from a Gaussian (non-Markovian) environment. Nevertheless, an environment described by this model does not induce a QDPT when the environment is only coupled to one site. This is reflected as a persistent minimum of the transmittance at $\varepsilon=0$ which can be attributed to the light-tails of the Gaussian distribution and specific energy correlations of the model. Therefore, no collapse of the transmittance is observed and both the bonding and antibonding energies seem to conserve their molecular identity. 
 
Finally, when both sites are under the action of an environment, either Lorentzian or Gaussian, the \textbf{through-atom} transmittance fades out as the coupling with the environment grows. Thus, transport becomes too small to manifest the QDPT that is known to exist for Lorentzian environments\cite{Pa07}. For the Gaussian environments one can observe how the renormalized transmittance still corresponds to the combination of two separated peaks becoming wider and moving away from each other. This may indicate that an accidental degeneracy in the Fock-space allows \(V_{AB}\) to dominate the small signal. This occurs  in spite of the physical wisdom that when each atom is strongly affected by an environment the bonding coupling  should not have a significant effect, as occurs for a Lorentzian environment. 
These results do not necessarily imply the absence of molecular dissociation but rather manifest that a steady-state through-atom tunneling measurement might  not always be an adequate tool to test it. 

In summary, we conclude that the transmittance spectrum (e.g. tunneling microscopy measurements) has dependencies on the specific model for the environment and on the interferences between the incoming and outgoing states. While it might not always reflect the energetic of the molecular dissociation/formation, it should not be discarded as tool that gives access to the resonances in the molecular spectrum, which, in turn, proved to quantify the stability of a given molecular configuration in the presence of a catalyst or other external degrees of freedom.

\appendix

\section{Fermi Golden Rule.}

In an ordered tight-binding linear chain with site energies \(E_d\) and nearest neighbors couplings \(V_d\), the infinite order perturbation theory can be fully solved within the Green's functions formalism\cite{PaFoMe02}. This solution is expressed in the self-energy that corrects a single orbital from the presence of the infinite orbitals at its right (left) which results from the Dyson equation:
\begin{equation}
\Sigma_d(\varepsilon)=\lim_{\eta\rightarrow\infty}\frac{V^{2}}{ \varepsilon-(E_{d}-\mathrm{i}\eta)-\Sigma_d(\varepsilon)}
=\Delta_d(\varepsilon)-\mathrm{i}\Gamma_d(\varepsilon).
\end{equation}
Similarly, the self-energy of an adatom B coupled to the surface of a semi-infinite chain with eigenstates $\left\vert  k\right\rangle $ through the hopping  \(V_0\) , the self energy results\cite{An78},
\begin{equation}
\Sigma_B(\varepsilon)=\lim_{\eta\rightarrow\infty}\frac{V_0^{2}}{ \varepsilon-(E_{d}-\mathrm{i}\eta)-\Sigma_d(\varepsilon)}\\
=\lim_{\eta\rightarrow0}\lim_{N\rightarrow\infty}
\sum_k^N    \frac{\left\vert
V_{0}\left\langle 1\right\vert \left. k\right\rangle  \right\vert ^2}{ \varepsilon-E_{k}+\mathrm{i}\eta} \\
=\left\vert \dfrac{V_{0} }{V_d}\right\vert ^{2} \Sigma_{d}(\varepsilon).
\end{equation}
Here, $\left\vert V_{0}\right\vert ^{2}$ is the second moment of the coupling of the orbital $B$ with the only substrate orbital with finite overlap. Through a Fourier transform in the \(\varepsilon\) variable one obtains all the Green's function time dependences, i.e. the retarded quantum propagators. This, provides all the subtleties of the quantum dynamics of the system which includes: 1) short time quadratic decay, 2) exponential decay for intermediate times. 3)power law decay for long times and 4) a survival collapse interference and the transition time between regimes 2) and 3) \cite{RfPa06, Pa07, Ga_Or21}.  
The simpler Fermi Golden Rule approximation results results by replacing the self-energy functions by constant values. For example,  the energy broadening $\Gamma_{B}$ produced in the adatom by its coupling with the metal is:
\begin{align}
\Gamma_{d} &  =\left.  \Gamma_{d}\left(  \varepsilon\right)  \right\vert
_{\varepsilon=E_{d}}=\left.  \lim_{\eta\rightarrow0}\int\mathrm{d}k\left\vert
V_{Bk}\right\vert ^{2}\frac{\eta}{\left(  \varepsilon-E_{k}\right)  ^{2}%
+\eta^{2}}\right\vert _{\varepsilon=E_{d}}\label{SE-Gamma}\\
&  =\left.  \pi\left\vert V_{0}\right\vert ^{2}N_{1}(\varepsilon)\right\vert
_{\varepsilon=E_{d}}\label{Gamma-FGR}\\
&  =\left.  \pi\left\vert V_{0}\right\vert ^{2}\frac{1}{\pi V^{2}}\sqrt
{V^{2}-\left(  \varepsilon-E_{s}\right)  ^{2}}\right\vert _{\varepsilon=E_{d}%
}\\
&  =\left\vert \dfrac{V_{0}}{V}\right\vert ^{2}\Gamma_{d}(\varepsilon
)_{\varepsilon=E_{d}},
\end{align}
where $N_{1}(\varepsilon)$ is the \textit{density of directly connected states}, i.e. the substrate energies on which this surface orbital decomposes. Similarly,
\begin{equation}
\Delta_{d}=\left.  \Delta_{d}\left(  \varepsilon\right)  \right\vert
_{\varepsilon=E_{d}}=\left.  P\int\mathrm{d}k\frac{\left\vert V_{Bk}
\right\vert ^{2}}{\varepsilon-E_{k}}\right\vert _{\varepsilon=E_{B}}.
\end{equation}

 Thus, the above procedure contains the thermodynamic limit of infinite number of orbitals in the substrate that enables the Fermi Golden Rule (FGR) approximation for the electron transfer dynamics. It is here where non-Hermiticity of the molecular effective Hamiltonian sneaks in. 

 Because of the regularization of the poles with the infinitesimal \(\rm i \eta\), usual in the calculus with complex variables, we obtained an  \textit{imaginary} component in the energy. Thus the broadening actually denotes the actual divergence of such \textit{perturbation} as the new eigenstates are orthogonal to the unperturbed one ones.  Indeed, while a convergent perturbation series yields real eigenenergies, which are \textit{isolated poles} in the real axis, a divergent series yields \textit{resonances}, \textit{i.e.} poles in the complex plane whose  imaginary part $\Gamma_{d}$ accounts for the mean life of the atomic orbital and whose approximate real part is the energy shift $\Delta_{A}$. 

\section{Asymmetric molecule and a environment.} \label{A:Asymmetric}

After the Lanczos redefinition of the substrate collective orbitals, if we consider the system composed of the two atoms of the molecule plus the first two (redefined) substrate collective orbitals, we arrive to the following  Hamiltonian:
\begin{equation}
    \mathbb{H}=\left(
\begin{array}{cccc}
 E_2 & -V_d & 0 & 0 \\
 -V_d & E_1 & -V_0 & 0 \\
 0 & -V_0 & E_B & -V_{AB} \\
 0 & 0 & -V_{AB} & E_A \\
\end{array}
\right),
\end{equation}
where we are interested in finding its ``eigenvalues'' or poles of the Green's function. For simplicity, let's consider $E_B=E_1$. A practical strategy to write an effective Hamiltonian between $2$ and $A$ is, first to rewrite $\mathbb{H}$ using the $1$-$B$ bonding-antibonding basis,
\begin{equation}
    \mathbb{H}'=\left(
\begin{array}{cccc}
 E_2 & -V_d/\sqrt{2} & V_d/\sqrt{2}  & 0 \\
 -V_d/\sqrt{2} &E_B -V_0 & 0 & -V_{AB}/\sqrt{2} \\
 V_d/\sqrt{2}  & 0 & E_B+V_0 & -V_{AB}/\sqrt{2} \\
 0 & -V_{AB}/\sqrt{2} & -V_{AB}/\sqrt{2} & E_A \\
\end{array}
\right)=\mathbb{H}'_0+\mathbb{V}',
\end{equation}
where $\mathbb{H}'_0$ denotes the diagonal part of the Hamiltonian, and $\mathbb{V}'$ the off-diagonal part. 

 Now we resort to a canonical transformation to convert the hopping through $b(1B)$ and $a(1B)$ in an effective hopping between $2$ and $A$. So, let's consider $\mathbb{U}=e^{\mathrm{i}\mathbb{W}}$, a canonical transformation matrix that diagonalizes $\mathbb{H}'$, that is, $\mathbb{H}_{\mathrm{eff.}}=\mathbb{U}^{-1}\mathbb{H}'\mathbb{U}$. Following the usual expansion for the effective Hamiltonian $\mathbb{H}_{\mathrm{eff.}}=\mathbb{H}'_0+\mathbb{V}'+\mathrm{i}([\mathbb{H}'_0,\mathbb{W}]+[\mathbb{V}',\mathbb{W}])+\frac{\mathrm{i}}{2}[[\mathbb{H}'_0,\mathbb{W}],\mathbb{W}]...$ the idea is to find $\mathbb{W}$ such that it removes the terms that are linear in the perturbation, i.e. $\mathbb{V}'+\mathrm{i}[\mathbb{H}'_0,\mathbb{W}]=0$, which yields:
\begin{equation}
    W_{nm}=-\mathrm{i}\frac{V'_{nm}}{E_m-E_n}
\end{equation}
for the elements of the transformation matrix $\mathbb{W}$. The transformed Hamiltonian (to quadratic order in $\mathbb{V}'$) is:
\begin{equation}
    \mathbb{H}_{\mathrm{eff.}}\approx \mathbb{H}'_0+\frac{\mathrm{i}}{2}[\mathbb{V}',\mathbb{W}],
\end{equation}
which in our case generates an effective Hamiltonian between $2$ and $A$:
\begin{equation}
  \mathbb{H}_{\mathrm{eff.}}(\varepsilon)= \left(
\begin{array}{cc}
 E_2+\Delta_2 & V_{A2} \\
 V_{A2} & E_A+\Delta_A\\
\end{array}
\right)
\end{equation}
where
\begin{eqnarray}
V_{A2}&=&-\frac{V_{AB} V_{d}}{4 (E_2-(E_B +V_0)}+\frac{V_{AB} V_{d}}{4
   (E_2-(E_B -V_0))}\nonumber\\
   &+&\frac{V_{AB} V_{d}}{4 (E_A-(E_B +V_0))}-\frac{V_{AB} V_{d}}{4
   (E_A-(E_B -V_0))}\label{A:EqV2A}\\
   &\approx& \frac{V_{AB}V_d}{V_0} \\
\Delta_2&=&-\frac{V_{d}^2}{2 (E_2-(E_B +V_0))}-\frac{V_{d}^2}{2 (E_2-(E_B -V_0))}\simeq 0\\
\Delta_A&=& \frac{V_{d}^2}{2 (E_A-(E_B +V_0))}+\frac{V_{d}^2}{2 (E_A-(E_B -V_0))}\simeq 0 \label{A:EqDeltaA}.
\end{eqnarray}
This terms are usual expressions for the energy in second-order perturbation theory and the contributions of the $B1$ bonding and antibonding states are easily recognizable. In the energy corrections, $\Delta_{2}$ and $\Delta_{A}$, those contributions affect the energy in opposite ways, canceling each other while in the hopping $V_{A2}$ both the bonding and antibonding states interfere constructively to give a \textit{positive} coupling \cite{LePaD90}. 

\begin{figure}
     \centering
     \includegraphics[width=0.5\textwidth]{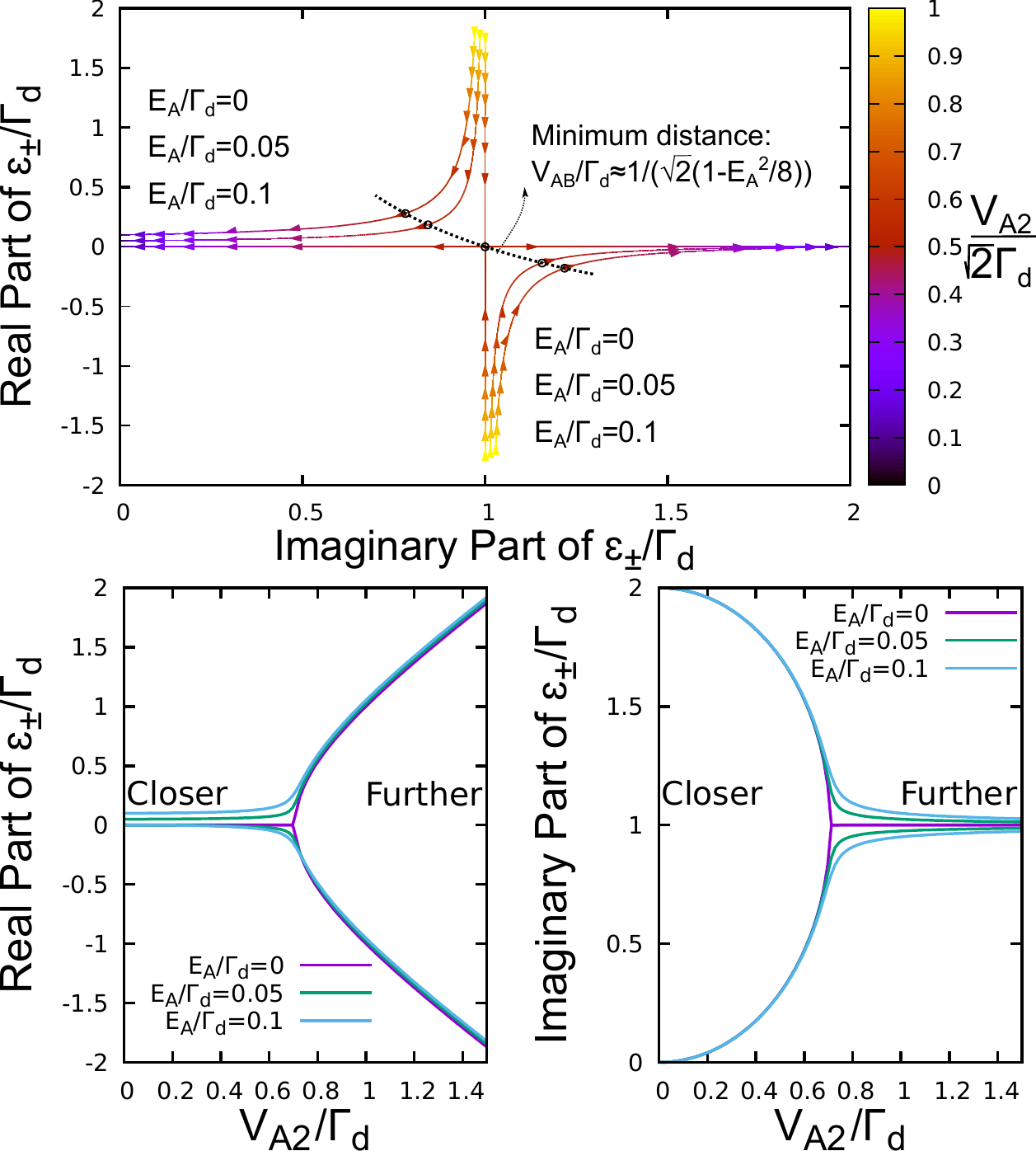}
     \caption{Top: Real and imaginary part of the resonances  $\varepsilon_{\pm}/\Gamma_d$ within the \(d\)-band for $E_B=0$ and three values of $E_A\geq 0$ as a parametric function of $V_{A2}/\Gamma_d$ around the critical point. Decreasing distances are indicated by the arrows. A \textit{positive} real parts represent \textit{antibonding} nature of the \textit{symmetric} resonant state while a \textit{negative} one represents a \textit{bonding} one. The imaginary part is the resonance broadening. We observe how for $E_A=0$ the curves cross at the exceptional point when $V_{A2}/\Gamma_d=\frac{1}{\sqrt{2}}$ . For $E_A\gtrsim  0$ there is an avoided crossing. The dashed black line pinpoint the minimum distances that generalize the critical values   $V_{A2}/\Gamma_d\simeq\frac{1}{\sqrt{2}}\frac{1}{1-(E_A-E_B)^2/8V_d^2}$. Bottom: Real and imaginary parts of  $\varepsilon_{\pm}/\Gamma_d$  as a function of $V_{A2}/\Gamma_d$, for $E_A=0$  at $V_{A2}/\Gamma_d=\frac{1}{\sqrt{2}}$ the curves collapse or bifurcate with decreasing distances, while for $E_A\gtrsim 0$ this criticality is only approximate.}
     \label{parametric}
 \end{figure}

Finally, as the substrate's orbital 2 is connected to the rest of the substrate's orbitals, a self energy $\Sigma_2(\varepsilon)$ should be introduced to the diagonal energy $E_2$. The energy dependence of the self energy $\Sigma_2(\varepsilon)=V_d^2/(\varepsilon-E_d-\Sigma(\varepsilon))=\Sigma(\varepsilon)$ should only be neglected in the imaginary part (where the corrections are one order higher), as the real part plays an important role in the modification of the poles. That means, $\Sigma(\varepsilon)\simeq \varepsilon/2-\mathrm{i}\Gamma_d=\varepsilon/2-\mathrm{i}V_d$. Subsequently, the equation for the resonant energies becomes:

\begin{equation}
  |\varepsilon\mathbb{I}-\widetilde{\mathbb{H}}_{\mathrm{eff.}}|  = \left|
\begin{array}{cc}
 \varepsilon -(E_2+ \varepsilon/2-\mathrm{i}\Gamma_d)& -V_{A2} \\
-V_{A2} & \varepsilon - E_A\\
\end{array}
\right|=0,
\end{equation}
where it follows,  
\begin{equation}
  \varepsilon_{\pm}=\frac{E_A}{2}+E_2-\mathrm{i} \Gamma_d \pm\sqrt{\left(-\mathrm{i} \Gamma_d -\frac{E_A}{2}-E_2\right)^2-2 \left(\mathrm{i} \Gamma_d  E_A+E_A
   E_2-V_{A2}^2\right)}
\end{equation}
for simplicity, let's take $E_2=0$. If $E_A$ is also zero, the imaginary part of the poles collapse for $V^c_{A2}=\frac{1}{\sqrt{2}}\Gamma_d\Rightarrow V_0^c=\sqrt{2}V_{AB}$ while the real part bifurcates, as it was discussed in the main text. When $E_A\neq 0$, the collapse is avoided, however, a chance in the behavior of the poles can be identify as it is illustrated in Fig. \ref{parametric}. To generalized this change in behavior we have studied the points where the distance between the two curves is minimum, for a small $E_A$ this yields 
\begin{equation}
V_{AB}/\Gamma_d=\frac{1}{\sqrt{2}}\frac{1}{1-\tfrac{E_A^2}{8V_d^2}} \Rightarrow V^c_0=\sqrt{2}V_{AB} (1-\tfrac{(E_A-E_B)^2}{8V_d^2}).
\end{equation} This means that an asymmetry on the molecule helps the isolation of $E_A$ although no non-analyticity is observed as it occurs in a non-accessible region of the parameters space.

The same results can be obtained from the now standard Recursive Green's Function Method\cite{ThKi81,LePaD90,PaMe01}. This uses the original basis by decimating the sites $1$ and $B$ and analyzing the poles of the Green's function  $\mathbb{G}(\varepsilon)=(\varepsilon \mathbb{I}-\widetilde{\mathbb{H}}_{\mathrm{eff.}}(\varepsilon))^{-1}$ .  Where
\begin{equation}
  \widetilde{\mathbb{H}}_{\mathrm{eff.} }(\varepsilon)= \left(
\begin{array}{cc}
 E_2+\Sigma_2(\varepsilon)+\Delta_2(\varepsilon) & V_{A2}(\varepsilon) \\
 V_{A2}(\varepsilon) & E_A+\Delta_A(\varepsilon) \\
\end{array}
\right),
\end{equation}
and the following approximations should be done:
\begin{eqnarray}
\Delta_2(\varepsilon) &\approx& \Delta_2(E_2)\\ 
\Delta_A(\varepsilon) &\approx& \Delta_A(E_A)\\ 
V_{A2}(\varepsilon) &\approx& (V_{A2}(E_2)+V_{A2}(E_A))/2.
\end{eqnarray}These magnitudes can be written in the form of the ones obtained through the canonical transformation (Eq. \ref{A:EqV2A}-\ref{A:EqDeltaA}) by using partial fraction decomposition.

\section{More on the Numerical through-bond transmittance with a Lorentzian environments in a configuration.}\label{A:TBLQ}

\begin{figure}
     \centering
     \includegraphics[width=0.5\textwidth]{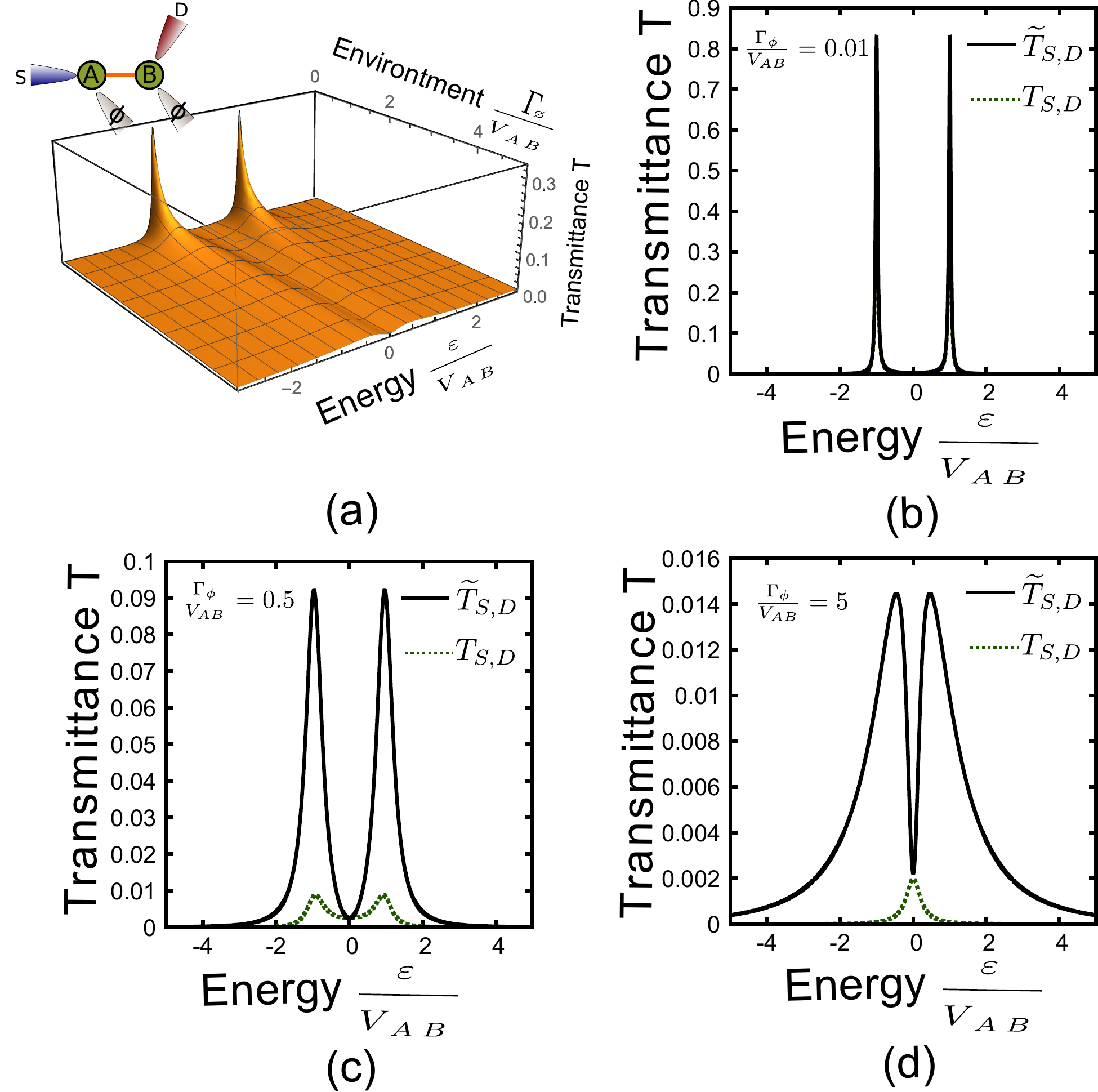}
     \caption{Top-Left: \textbf{Through-bond} effective transmittance as a function of the energy and the environment strength $\Gamma_{\phi}$. Right panel and Bottom-Left figures: Effective (black-solid lines) and coherent (direct, green-dashed lines) transmittance as a function of the energy for $\Gamma_{\phi}/V_{AB}={0.01,0.5,5}$.}
     \label{TBLQ}
 \end{figure}

In the results section it was discussed how, for symmetric Lorentzian environment in a through-bond tunneling, an antiresonance at the $\varepsilon=0$ appears. In this appendix section expand on it by presenting plotting the total effective transmittance $\widetilde{T}_{S,D}$ along with the coherent transmittance $T_{S,D}$. In Fig. \ref{TBLQ}-a we observe a similar behavior of the transmittance than in the Gaussian environment, i.e. there is a minimum at the center and the transmittance decreases as $\Gamma_{\phi}$ increases. However, although there is a minimum at the center, the transmittance is not zero and the analytical expression allow us to identify this effect as a product of the coherent (direct) transmittance. Furthermore, the antiresonance is a product of the destructive interference of electrons affected by the environment in this specific leads configuration. In fig. \ref{TBLQ}-b to \ref{TBLQ}-d we observe how, as $\Gamma_{\phi}$ increases, the transmittance change from being practically a direct transmittance for $\Gamma_{\phi}/V_{AB}=0.001$ to be principally defined by the incoherent term for  $\Gamma_{\phi}/V_{AB}=5$. As can be noted in this sequence, the direct term $T_{S,D}$ have a collapse of the two peaks for a finite $\Gamma_{\phi}$.

\section{Analysis of the Lorentzian and Gaussian Environments.}

In this section, we explore analytically the transmittance in the system by assuming $N\rightarrow\infty$ and summing (averaging) the transmittances with their probability. In order to have general result the wide band approximation will be omitted. Therefore, $\Sigma(\varepsilon)=\Delta(\varepsilon)-\rm{i}\Gamma_l(\varepsilon)=\frac{\varepsilon}{2}-\rm{i}V\sqrt{1-\frac{\varepsilon^2}{4V^2}}$ accounts for the self energies of the leads.
  
\subsection{The Lorentzian environment.}\label{A:LorEnv}
  
Let's consider a two-site molecule, with source (in) and drain (out) leads on one site and a Lorentzian environment on the other. This environment, is supposed to affect the system in the same way that the Gaussian environment, i.e. by introducing a sum of transmittance over Lorentzian distributed energies. The Green Function is given by,
\begin{equation}
\label{GreenFc-AA}
G_{LR}(\varepsilon)=\frac{1}{\varepsilon-E_0-\frac{V_{AB}^2}{\varepsilon-E_i}-2\frac{V_l^2}{V^2} \Sigma(\varepsilon)},
\end{equation} where $V_l$ is the hopping with the leads, and $\Sigma(\varepsilon)$ its self energy, and, $E_i$ a Lorentzian random variable. 

Using the properties of Lorentzian random variables (rv) \cite{Egg74} and maintaining his notation, where $E_i\rightarrow (0,\Gamma_\phi)$ means that $E_i$ is a Lorentzian random variable with a probability density function given by $P(E_i)=\frac{\Gamma_\phi}{\pi}\frac{1}{\Gamma^2_\phi+E_i^2}$. Then, 

\begin{eqnarray}
    E_i&\rightarrow& (0,\Gamma_\phi)\implies E_{i,\varepsilon}=\varepsilon-E_i\rightarrow(\varepsilon,\Gamma_\phi)\implies\\
   E'_{\varepsilon,\Gamma_\phi}=\frac{-V_{AB}^2}{E_{i\varepsilon}}&\rightarrow& (\frac{-V_{AB}^2\varepsilon}{\varepsilon^2+\Gamma_\phi^2},\frac{V_{AB}^2\Gamma_\phi}{\varepsilon^2+\Gamma_\phi^2}),
\end{eqnarray} 
so until here, and setting $E_0=0$, we have:
\begin{eqnarray}
    G_{LR}(\varepsilon)&=&\frac{1}{\varepsilon-E'_{\varepsilon,\Gamma_\phi}-2\frac{V_l^2}{V^2} \Sigma(\varepsilon)}\\=
    &=&\frac{1}{\varepsilon-E'_{\varepsilon,\Gamma_\phi}-2\frac{V_l^2}{V^2}(\varepsilon/2-\mathrm{i}V\sqrt{1-\varepsilon^2/(4V^2)})}\\
    &=&\frac{1}{(\varepsilon-\frac{V_l^2}{V^2}\varepsilon)-E'_{\varepsilon,\Gamma_\phi}+\mathrm{i}2\frac{V_l^2}{V}\sqrt{1-\varepsilon^2/(4V^2)})}.
\end{eqnarray}
Thinking $(\varepsilon-\frac{V_l^2}{V^2}\varepsilon)$ as a non-random variable which only modify the mean value of $E'_{\varepsilon,\Gamma_\phi}$, maintaining  $\mathrm{i}2\Gamma_l(\varepsilon)=\mathrm{i}2\frac{V_l^2}{V}\sqrt{1-\varepsilon^2/(4V^2)})$ as an imaginary number, and calculating the squared absolute value of the Green function we have:
\begin{eqnarray}
    |G_{LR}(\varepsilon)|^2&=&|\frac{1}{E''_{\varepsilon,\Gamma_\phi}+\rm{i}\Gamma_l(\varepsilon))}|^2\\
    &=&\frac{1}{(E''_{\varepsilon,\Gamma_\phi})^2+(2\Gamma_l(\varepsilon))^2},
\end{eqnarray}
where $E''_{\varepsilon,\Gamma_\phi}$ is a Lorentzian random variable of parameters $x_0=(\varepsilon-\frac{V_l^2}{V^2}\varepsilon)-\frac{V_{AB}^2\varepsilon}{\varepsilon^2+\Gamma_\phi^2}$ and $\gamma=\frac{V_{AB}^2\Gamma_\phi}{\varepsilon^2+\Gamma_\phi^2}$ (from now on the sub-indices ${\varepsilon,\Gamma_\phi}$ will be omitted). Therefore, the averaged Green function is given by:
\begin{eqnarray}
\langle |G_{LR}(\varepsilon)|^2\rangle&=&\int_{-\infty}^{\infty} P(E_i)|G_{LR}(\varepsilon)|^2dE_i\\
&=&\int_{-\infty}^{\infty} P(E'')\frac{1}{(E'')^2+(2\Gamma_l(\varepsilon))^2}dE''\\
&=&\int_{-\infty}^{\infty} \frac{\gamma}{\pi}\frac{1}{(E''-x_0)^2+\gamma^2}\frac{1}{(E'')^2+(2\Gamma_l(\varepsilon))^2}dE''.
\end{eqnarray}
The principal value of the integral above is:
\begin{eqnarray}
\langle |G_{LR}(\varepsilon)|^2\rangle&=&\frac{1+\frac{\gamma}{2\Gamma_l(\varepsilon)}}{x_0^2+(2\Gamma_l(\varepsilon)+\gamma)}\\
&=&\frac{1}{x_0^2+(2\Gamma_l(\varepsilon)+\gamma)}+\frac{\frac{\gamma}{2\Gamma_l(\varepsilon)}}{x_0^2+(2\Gamma_l(\varepsilon)+\gamma)}\label{LorSta-GF}\\
&=&
\frac{1+\frac{V_{AB}^2\Gamma_\phi}{(\Gamma_\phi^2+\varepsilon^2)2\Gamma_l(\varepsilon)}}{(\varepsilon-V_l^2\varepsilon-\frac{V_{AB}^2\varepsilon}{\Gamma_\phi^2 + \varepsilon^2})^2 + (\frac{V_{AB}^2\Gamma_\phi}{\Gamma_\phi^2+\varepsilon^2} + 2\Gamma_l(\varepsilon))^2}.
\end{eqnarray}

The generalization to a molecule with Lorentzian environment in the two sites is straight forward: In this case, $E''_{\varepsilon,\Gamma_\phi}$ is also a Lorentzian random variable, but with parameters $x_0=(\varepsilon-V_l^2\varepsilon)-\frac{V_{AB}^2\varepsilon}{\varepsilon^2+\Gamma_\phi^2}$ which remains the same and $\gamma=\frac{V_{AB}^2\Gamma_\phi}{\varepsilon^2+\Gamma_\phi^2}+\Gamma_\phi$ which have an extra $\Gamma_\phi$. This extra $\Gamma_\phi$ is responsible of the broadening of the transmittance for large environments.

These solutions match perfectly the D'Amato-Pastawski solutions. Furthermore, multiplying \eqref{LorSta-GF} by $4\Gamma(\varepsilon)^2$ we can identify the first term of the direct transmittance and the second term is the indirect transmittance.

Finally, when the transport is through the bond with a asymmetric environment, the equivalence of the D'Amato-Pastawski model with this Lorentzian environment fluctuation seems to be valid (numerically). Although, an analytic analysis has not been made. For symmetric environments in through-bond transport, this relation do not holds.

 \subsection{The Gaussian environment.}\label{A:GaussEnv}
 
 A similar analysis, with some extra difficulties, can be done with the Gaussian environment. In this case, the minimum of the transmittance observed at $\varepsilon=0$ can be understood. As with the Lorentz random variables, a sum of two Gaussian random variables or the multiplication of a Gaussian RV by an scalar give as result another Gaussian random variable but with some differences, for example, for Gaussians the squares of the widths are added $\Sigma_T^2=\Sigma_1^2+\Sigma_2^2$. 
 
 Consequently, $\varepsilon-E_i$ in the Green function \eqref{GreenFc-AA}, can be thought as a Gaussian RV, just centered at $\varepsilon$. The most fundamental difference with the treatment we did for Lorentzian environments is that the inverse of a Gaussian RV is not a Gaussian RV. Furthermore, they are quite different, the inverse distribution is bimodal, and the first and higher-order moments do not exist. However, for real-valued shifted reciprocal function, these moments do exist in a principal value sense. Concretely, if $X$ if our Gaussian RV, with a probability density function (PDF) $g(x,x_0,\sigma)=\frac{1}{\sigma\sqrt{2\pi}}e^{-\frac{(x-x_0)^2}{2\sigma^2}}$ the PDF of $1/X$ is:
 \begin{equation}
    f(y,x_0,\sigma)=\frac{1}{y^2}g(1/y,x_0,\sigma)=\frac{1}{y^2\sigma\sqrt{2\pi}}e^{-\frac{(1/y-x_0)^2}{2\sigma^2}}.
 \end{equation}
As we noted before, this PDF is bimodal with a minimum of probability at the center ($y=0$), this is a direct consequence of the light tail of the Gaussian, which generated very high numbers with an ``more than exponential'' decreasing probability. In opposition, $f$ has Lorentz tails, a consequence of the higher, but finite, probability of having $X$ near zero \cite{Leh12}.

 \begin{figure}
     \centering
     \includegraphics[width=0.5\textwidth]{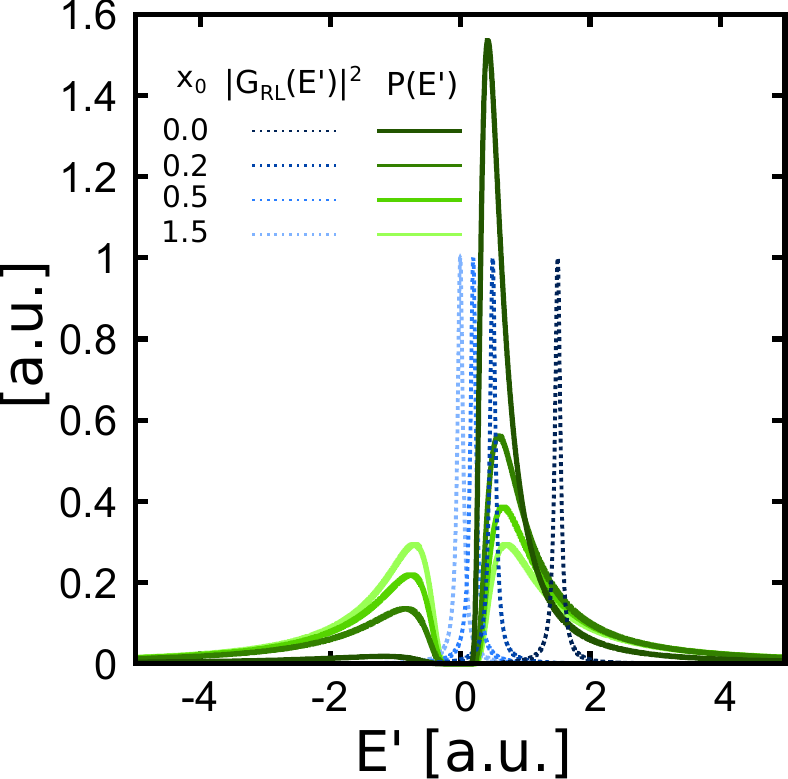}
     \caption{Schematic representation (in arbitrary units) of $P(E')$ (solid Green lines) and  $|G_{RL}(E',\varepsilon)|^2$ (dotted blue lines). As the parameter $x_0$ increases (darker lines). We observe how as $|G_{RL}(E',\varepsilon)|^2$ displaces to the right and the probability distribution $P(E')$ accumulated in the positive region of $E'$ increasing the transmittance as $\varepsilon$ becomes non-zero.}
     \label{InverseGauss}
 \end{figure}
 
This properties affects directly the transmittance in systems with Gaussian environment. As we pointed out, the minimum of $f$ at the center is responsible of the minimum observed in the transmittance. To observe this effect is useful to write the average of the Green function as follows:
 
 \begin{eqnarray}
 \langle |G_{LR}(\varepsilon)|^2\rangle&=&\int_{-\infty}^{\infty} P(E_i)|G_{LR}(\varepsilon)|^2dE_i\\
&=&\int_{-\infty}^{\infty} P(E')\frac{1}{(\varepsilon-\hat{E_A}-E')^2+(2\Gamma_l(\varepsilon))^2}dE''\\
&=&\int_{-\infty}^{\infty} P(E')|G_{RL}(E',\varepsilon)|^2dE'',
 \end{eqnarray}
 where $E'$ is a RV with a PDF $P(E')=\frac{1}{E'^2}g(1/E',x',\sigma')$ where $x'=\varepsilon$ and $\sigma'=\sigma$. For $\varepsilon=0$, $P(E')$ is symmetric and $|G_{RL}(E',\varepsilon)|^2$ is a Lorentzian function centered at zero, in this case, the minimum of $P(E')$  coincides with the maximum of $|G_{RL}(E',\varepsilon)|^2$, lightening its weight on the integral. As $\varepsilon$ becomes different of zero, but still small, the symmetry in $P(E')$ is broken and one of the modes gain probability, simultaneously, $|G_{RL}(E',\varepsilon)|^2$ moves in the direction of the preferred mode, increasing the overlap, the value of the integral, and therefore, the Magnitude of the transmittance. Finally, for bigger $\varepsilon$, the Lorentzian pass through the mode, and it is centered at the tail of $P(E')$, reducing again the value of the transmittance. This effects can be observed in figure \ref{InverseGauss}.
 
 Finally, the generalization to a molecule with two environment is similar to the one developed to the Lorentzian environment but not so clean. In this case, we will have a sum of two RV, one with a Gaussian distribution, and the other with an inverse Gaussian distribution. For weak environments, i.e. for small $\sigma$, and specially for small Fermi energies $\varepsilon$, the inverse Gaussian distribution dominates, generating the two peaks transmittance. However when as $\sigma$ increases the minimum at the center of the distribution disappears, the general behavior is therefore a consequence of the first site and its environment as in its Lorentzian partner.

\bibliographystyle{apalike}
%para referencias en forma de numeros:
%\bibliographystyle{ieeetr}

\bibliography{MQC-OTOC-NMR}
 
\end{document}